\documentclass[12pt,preprint,flushrt]{aastex}
\newcommand{\beq}{\begin{equation}}
\newcommand{\eeq}{\end{equation}}
\newcommand{\e}{$^{-1}$}

\newcommand{\caln}{{\cal N}}
\newcommand{\ee}{$^{-2}$}

\newcommand{\ag}{\mbox{ \raisebox{-.4ex}{$\stackrel{\textstyle >}{\sim}$} }}
\newcommand{\al}{\mbox{ \raisebox{-.4ex}{$\stackrel{\textstyle <}{\sim}$} }}
\newcommand{\co}{{\rm CO}}
\newcommand{\davdg}{\Delta A_{V,\,\rm DG}}
\newcommand{\dnsdg}{\Delta N_{\rm DG}}

\newcommand{\fdark}{f_{\rm DG}}
\newcommand{\fv}{f_{\rm V}}

\newcommand{\hone}{\ion{H}{1}}
\newcommand{\htwo}{H$_2$}

\newcommand{\krho}{{k_\rho}}
\newcommand{\muh}{\mu_{\rm H}}

\newcommand{\pmax}{P_{\rm max}}
\newcommand{\pmin}{P_{\rm min}}

\newcommand{\pth}{P_{\rm th}}
\newcommand{\pthtp}{P_{\rm th}^{\rm 2p}}
\newcommand{\pthturb}{P_{\rm th}^{\rm turb}}

\newcommand{\rco}{R_{\co}}
\newcommand{\rht}{R_{\rm H_2}}
\newcommand{\rtot}{R_{\rm tot}}

\newcommand{\vecr}{{\bf r}}
\newcommand{\xc}{x_{\rm C}}
\newcommand{\xcp}{x_{\rm C^+}}
\newcommand{\xco}{x_{\co}}

\newcommand{\xo}{x_{\rm O}}
\newcommand{\xoh}{x_{\rm OH}}

\slugcomment{Accepted for Publication in the Astrophysical Journal}

\shortauthors{Wolfire et al.}
\shorttitle{The Galactic ISM}
\received{}
\begin{document}

\title{The Dark  Molecular Gas}

\author{
Mark\ G. Wolfire}
\affil{Department of Astronomy, University of Maryland,
College Park, MD 20742-2421}
\email{mwolfire@astro.umd.edu}

\author{
David Hollenbach}
\affil{SETI Institute, 515 N. Whisman Road, 
Mountain View, CA 94043}
\email{dhollenbach@seti.org}

\and

\author{
Christopher F.\ McKee}
\affil{Physics Department and Astronomy Department,
University of California at Berkeley, Berkeley, CA 94720}
\email{cmckee@astron.berkeley.edu}

\begin{abstract}

The mass of molecular gas in an interstellar cloud is often measured using line emission
from low rotational levels of CO, which are sensitive to the CO mass, and then
scaling to the assumed molecular hydrogen H$_2$ mass. However, a significant H$_2$ mass
may lie outside the CO region, in the outer regions of the molecular cloud
 where the gas phase carbon resides in C or C$^+$.
Here, H$_2$ self-shields or is shielded by dust from UV photodissociation, whereas CO
is photodissociated.  This H$_2$ gas is ``dark" in molecular transitions because
of the absence of CO and other trace molecules, and because H$_2$ emits so weakly
at temperatures 10 K $< T < 100$ K typical of this molecular component.
This component has been indirectly observed through other tracers of mass such
as gamma rays produced in cosmic ray collisions with the gas 
and far-infrared/submillimeter wavelength dust continuum radiation.
In this paper we theoretically model this dark mass and find that the fraction of the 
molecular mass in this dark component is remarkably constant ($\sim 0.3$ for average visual extinction
through the cloud $\bar A_V \simeq 8$) and insensitive
to the incident ultraviolet radiation field strength, the internal density distribution, and the mass
of the molecular cloud as long as $\bar A_V$, or equivalently, the product of the average
hydrogen nucleus column and the metallicity through the cloud, is constant.   We also find that the dark mass
fraction increases with decreasing $\bar A_V$, since relatively more molecular H$_2$
material lies outside the CO region in this case.

\end{abstract}


\section{INTRODUCTION}
\label{sec:intro}

Various observations have indicated that a substantial amount
of interstellar gas exists in the form of molecular hydrogen (${\rm H_2}$) along
with ionized carbon (${\rm C^+}$), but little or no carbon monoxide (CO).
The total mass in molecular hydrogen has been estimated from gamma ray
observations from COS-B \citep{bloemen1986} and EGRET (Energetic Gamma
Ray Experiment Telescope)
\citep{strong1996} and 
analysis of this data showed more gas mass 
than can  be accounted for in \ion{H}{1} and CO alone
\citep{grenier2005}. In addition, the dust column density maps of the Galaxy
from DIRBE, and maps of the 2MASS J-K extinction show additional gas not seen 
in \ion{H}{1} or CO \citep{grenier2005} and is
presumably molecular hydrogen. In molecular line observations 
and modeling of low column density molecular clouds the ${\rm H_2}/{\rm CO}$
ratio is found to be variable and much larger than 
 $10^{4}$ so that only a small fraction of
the C is in CO with the remainder presumably in ${\rm C^+}$
\citep{hollenbach2009,goldsmith2008}. 
\cite{reach1994} using infrared continuum maps of diffuse clouds
along with \ion{H}{1} and CO observations, found  
${\rm H_2}$ masses comparable to the \ion{H}{1} masses,
with only small amounts of CO. 
A mixture of ${\rm H_2}$ and ${\rm C^+}$
is also inferred to exist in diffuse clouds where the ${\rm H_2}$ and
CO columns are measured by UV absorption spectroscopy and the CO 
accounts for only a trace amount of C
\citep{sonnentrucker2007,burgh2007,sheffer2008}.
This 
component of the interstellar medium (ISM), termed ``dark gas'' by \citet{grenier2005},
is also inferred
to exist in extragalactic observations comparing far-infrared and CO mass
estimates especially in low metallicity galaxies \citep{israel1997,leroy2007}. 

Such an ${\rm H_2}$ and ${\rm C^+}$ layer is also predicted from
theoretical models of diffuse gas \citep{vandishoeck1988} and 
surfaces of molecular clouds \citep{tielens1985} that indicate 
the transition from ${\rm C^+}$ 
to CO is deeper into the cloud than the transition from H to ${\rm H_2}$.
Essentially, the theoretical models show that  H$_2$ self-shields itself
from UV photodissociation more effectively than CO.  This layer is ``dark" in rotational
H$_2$ transitions primarily because the ground state transition 0-0 S(0)
at 28 $\mu$m lies about $\Delta E/k \simeq 512$ K above ground; at the
temperatures 10 K $< T < 100$ K typical of the dark component the fluxes from the
H$_2$ rotational transitions lie below current sensitivities. This layer
is dark in CO because of its very low abundance.
 Although the dark gas layer is ``dark'' in H$_2$ and CO, it does emit mainly in 
[\ion{C}{2}] 158 $\mu$m fine-structure line emission.
In preliminary estimates (M.\ Wolfire et al.\ 2010, in preparation) 
the calculated 
local Galactic [\ion{C}{2}] emission from the WIM and CNM
(diffuse phases that are not associated with molecular clouds)
underestimate the COBE observations of the line emission in the 
plane by a factor of 1/3 to 1/2 and could be another indicator of dark gas. 
Similarly, \cite{shibai1991} and \cite{cubick2008} found the bulk of 
the [\ion{C}{2}] emission in the
Galactic plane seen by BIRT and COBE arises in neutral gas associated 
with molecular clouds. 

In this paper we present models of molecular cloud surfaces to 
estimate the mass of gas in the ``dark'' component. In \S 2 we discuss modifications
to existing photodissociation region (PDR) codes, and the modeling procedure.
In \S 3 we define the dark gas mass fraction and other parameters used
in our models, discuss the average gas density distribution assumed in our
modeling, and derive an expression for the dark gas fraction in terms
of parameters found in our numerical modeling procedure.  The
modeling results 
start in \S 4 with a simple isobaric cloud model in the limit 
of clumps that are optically thin to the incident radiation
field, resulting in an estimate of the dark gas fraction for a typical 
giant molecular cloud.  We then in \S 5 enhance this model with the inclusion of
turbulent pressure and find the dark gas fraction as a function
of incident field strength and cloud mass. We also discuss the
variation in the dark gas fraction as a function of metallicity (over a limited range),
the average hydrogen column through a cloud, the average visual extinction through a cloud,
and the opacity of the clumps. 
We compare our results with observations.  Finally, in \S 6 we conclude with a
discussion and summary.

\section{Models}
\label{sec:Models}

Our goal is to determine the mass in the
${\rm C^+}/{\rm H_2}$ layer---the dark gas---in molecular
clouds. Our analysis is based in part on the results
of a modified version of the photodissociation
region (PDR) code of \cite{kaufman2006}. 
In applying this code, we are making four main approximations:
First, we assume that the inhomogeneities in actual molecular clouds
can be approximated as clumps of density $n_c$ occupying a fraction
$\fv$ of the volume; we neglect the interclump medium, so that
the locally volume-averaged density, $\bar n$ is given by
$\bar n=\fv n_c$. Second, we assume that these clumps are
optically thin to UV radiation; as a result, the radiative transfer is the same as that
in a homogeneous medium with the mean density, $\bar n$.
The validity of this assumption is discussed in \S \ref{sec:justif} below, where we
show that even in the limit of very optically thick clumps, our models give
essentially the same dark gas mass fraction.
Third, in the PDR solution to the variation with depth of chemical abundances
and gas temperature, we assume that the layer of dark gas is not geometrically thick,
so that we can apply the results of a 1D slab model to the spherical problem.  However,
in calculating the mass of gas in the dark gas layer, we do take into account
the spherical geometry. Fourth, we assume that we can calculate the
chemical abundances in steady state. This assumption is discussed 
in \S \ref{sec:justifchem} after we have derived densities, turbulent
speeds and characteristic distances.
With these approximations, the physical conditions 
and chemical abundances within the cloud are a function of 
the optical depth from the cloud surface and we can
apply a single, one-dimensional PDR model to a continuous density distribution
in which the gas density $n_c$ inside the clumps can vary as a function
 of the distance $r$ of a clump from the center of the GMC.

The PDR models find the gas temperature in thermal balance
and the chemical equilibrium abundance of all dominant atomic
and molecular species as a function of depth into a gas layer.
The models require the incident 
far-ultraviolet (FUV; 6 eV $< h\nu < 13.6$ eV) radiation, extreme
ultraviolet (EUV, 13.6 eV $ < h\nu \la 100$ eV) radiation, soft
X-ray (100 eV $ \la h\nu \la 1$ keV)  
radiation, cosmic-ray flux, 
and either the spatial distribution in density or the spatial distribution in
thermal pressure.
If the density distribution is provided, the code will iterate
on the gas temperature at fixed density until thermal 
balance is achieved. If the 
thermal pressure 
($P_{\rm th}/k=n_tT$, where $n_t$ is the total particle density)
is provided, the code will iterate on both the
density and temperature.
With these inputs, 
the temperature, and the atomic and molecular abundances are
calculated self-consistently.

\subsection{Modified PDR Code}

The PDR code of \cite{kaufman2006} is based on two main parts. The first
part is derived from \cite{kaufman1999}, \cite{wolfire1990}, and
\cite{tielens1985} models and is the main program for calculating the
chemistry, thermal balance, and line emission. The second part is based
on the \cite{lepetit2006} Meudon PDR code and is used for all of the
molecular hydrogen processes. The \cite{kaufman2006} code
assumes a 1D UV  flux normal to the cloud surface.
Here we consider the average radiation field produced by all OB associations 
illuminating the giant molecular cloud (GMC) and the field at the cloud
surface
is expected to be relatively isotropic over $2\pi$ steradians. In order to account for the isotropic 
field, we assume a 1D flux incident at an angle of 
60 degrees to the normal. This unidirectional field has the same 
flux normal to the cloud surface as the isotropic field, and thus the same
energy input to the cloud, and it more closely approximates the depth 
penetration
of the field. The photo rates are modified to account 
for twice the path length of the incident field for a given $A_V$ 
measured normal to the surface. In the Meudon portion of the code, the 
isotropic field
is accounted for explicitly in the depth dependence of the photo rates.

We have also included several changes in the elemental abundances 
and chemical network that affect the carbon chemistry. 
We adopt the gas phase carbon abundance ${\cal A}_{\rm C}=1.6\times 10^{-4}$
from \cite{sofia2004},
where ${\cal A}_{\rm C} = n_{\rm C}/n$ and where $n_{\rm C}$ is the gas
phase number density of elemental carbon and $n$ is the number density 
of H nuclei.  We also include the ${\rm C^+}$ radiative and
dielectronic recombination rates from 
\cite{badnell2006a} and 
(Badnell 2006b)\footnote{http://amdpp.phys.strath.ac.uk/tamoc/DR},
who have found that the rates at low temperatures 
can be a factor of 2-3 higher than was used in the past
\citep[e.g.,][]{nahar1997}.  
Another change in the chemical network is to include OH
production on grains from \cite{hollenbach2009}. We adopt the 
fastest possible rate, having each collision  of O with a
grain combine with H and produce OH. This limit tends to 
maximize CO production and minimize the dark gas fraction. 
We also include the ion-dipole reaction rates 
for ${\rm C^+}$, ${\rm H_3^+}$, and ${\rm He^+}$,
reacting with OH and ${\rm H_2O}$
as discussed in
\cite{hollenbach2009}.

\subsection{External Radiation Field}
\label{subsec:ExternalRadiationField}

The external 
fluxes of FUV radiation, EUV/soft X-ray radiation, and cosmic rays
are model inputs and must be specified.
Associations produce bright
photodissociation regions on nearby cloud surfaces.  For example, 
the Trapezium cluster in Orion produces
the bright Orion PDR on cloud surfaces only $\sim 0.1$ pc distant.
As an association evolves, the Str\"omgren  region expands,
the embedded \ion{H}{2} region breaks out of the cloud, the resultant
champagne flow and blister evaporation disperses the cloud
and the distance between
the cloud and  association increases, thus allowing more of the
cloud to be illuminated than the initial hot spot.
The combination of cloud and association evolution,   
the distribution of associations around the cloud, and the continuous 
formation of new associations, elevates the average radiation
field illuminating the cloud to a field strength 
that is greater than the average interstellar radiation field.  
A detailed evaluation of this process is
presented in Wolfire, Hollenbach, \& McKee (2010, in preparation). 
They find the average field 
incident on clouds is on the order of $\sim 20$ times higher 
than the local Galactic interstellar field. 
We adopt the \cite{draine1978} 
interstellar radiation field that is equal to $\sim 1.7$ times the
\cite{habing1968} local interstellar field 
($1.6 \times 10^{-3}$ erg cm$^{-2}$ s$^{-1}$ for FUV photons). 
We also use the notation that 
$G_0$ is the radiation field in units of the Habing field and 
$G_0'$ is the field relative to the local Galactic field 
found by \citet{draine1978},
so that $G_0'= G_0/1.7$. 
To clarify the definition of $G_0'$, which is consistent with the original
definition
given in \cite{tielens1985}, $G_0'$ is the ratio of the photorates
at the surface
($A_V=0$) of our cloud to the same rates in the Draine field in free space.
Thus,
$G_0'= (2 \pi I)/(4 \pi I_D)=0.5I/I_D$, where $I$ is our incident isotropic
intensity and
$I_D= 2.2\times 10^{-4}$ erg cm$^{-2}$ s$^{-1}$ sr$^{-1}$ is the Draine
intensity for
the local ISM.\footnote{Note that an opaque cloud in the local ISRF
therefore experiences
$G_0'=0.5$, since the molecule at the surface is only illuminated from $2
\pi$ sterradians.}
Here we examine a range of FUV fields, 
$3< G_0' < 30$, consistent with (and generously encompassing)
the results of Wolfire et al.\ (2010). Throughout this paper
we simply scale the flux by $G_0'$ and keep the spectral 
energy distribution fixed as given by the \cite{draine1978}
field.

As discussed in \cite{wolfire03}, the 
soft X-ray/EUV flux from the diffuse interstellar 
field can be produced by a combination of supernova remnants
and stellar sources. The field shining on the molecular cloud
can be enhanced above the interstellar flux 
due to the OB associations near the molecular cloud. 
We assume that we are outside
any of the \ion{H}{2}\  regions associated
with the OB associations, and simply
scale the soft X-ray/EUV flux by the same factor
that we do for the effective far-ultraviolet radiation field 
$\zeta'_{{\rm XR}}= G_0'$.
A corollary of this assumption is that the fraction of
gas that is ionized is negligible.
\cite{wolfire03} adopted a standard soft X-ray/EUV
shielding column of neutral gas of column density 
$N = 1\times 10^{19}$ ${\rm cm^{-2}}$. 
The layer $N < 10^{19}$ ${\rm cm^{-2}}$ mainly consists of 
WNM gas photoevaporated from the molecular and CNM clump surfaces
and does not provide significant shielding to the FUV photodissociating
radiation field. We are mainly concerned with the deeper layers and 
assume that the clouds are shielded by a column of this order.
As we shall demonstrate in 
\S \ref{subsec:OpticallyThinP2}, 
the absorption of X-ray/EUV radiation
at columns in the range $10^{19}$ cm$^{-2}$ $<N \sim 10^{21}$ ${\rm cm^{-2}}$ leads to  
a drop in the thermal 
pressure by a factor of $\sim 10$
in the \hone\ layer in the cloud.

The cosmic-ray flux is also associated with
massive star evolution, but is not as likely to be
simply proportional to $G_0'$ as for the
EUV/soft X-ray flux. Since the temperatures of cloud interiors 
do not exceed $T=10-20$ K, the cosmic ray flux in cloud
interiors cannot 
be higher than a factor of $\sim 2-3$ times the diffuse ISM 
value or else the heating rate would push internal cloud
temperatures higher than observed. 
In most of our calculations, we therefore assume that cosmic-ray rates are not enhanced by the 
nearby molecular cloud OB associations. 

The cosmic ray rate can also affect the formation of CO in the diffuse
gas and outer molecular cloud layers through the production of OH
as discussed by \cite{vandishoeck1986,vandishoeck1988}. 
There are several observations, mainly of ${\rm H_3^+}$, that
indicate higher cosmic ray ionization rates by a factor of 
$\sim 10$ along select lines of
sight in the diffuse ISM \citep{indriolo2007}; however the
${\rm H_3^+}$ observations do not indicate higher 
cosmic ray rates in the denser, molecular clouds considered here
\citep{mccall1999}.
We adopt our standard rate for the majority of this paper,
but consider the effects of
higher cosmic ray rates in the outer portions of the cloud
in \S \ref{subsubsec:varcr},

\section{Density Distribution and Dark-Gas Fraction}
\label{sec:DensityDistribution}

To estimate the dark-gas fraction in a molecular cloud, 
we assume that it is spherical.
Let $\rco$ be the radius of the CO part of the cloud, defined by the condition
that the optical depth from $\rco$ to the surface in the $J=1-0$
transition is $\tau_{\co}=1$. Let $\rht>\rco$ be the radius of the part of the cloud
in which hydrogen is molecular; we define this
as the radius at which the mass density in \htwo\ molecules
equals that in H atoms. Finally, the total cloud radius, $\rtot$,
includes the \hone\ shielding layer that absorbs the FUV radiation incident upon
the cloud, enabling the gas to become molecular. 
The corresponding masses are the CO cloud mass, $M(\rco)$, 
the total molecular mass, $M(\rht)$, and the total cloud mass, $M(\rtot)$, respectively,
including the associated helium.
The dark gas extends from $\rco$ to the
atomic-molecular transition radius at $\rht$. 
The dark-gas mass fraction is then
\beq
f_{\rm DG} = \frac{M(\rht)-M(\rco)}{M(\rht)}.
\label{eq:fDG}
\eeq
Note that $f_{\rm DG}$ is the fraction of {\it molecular} (H$_2$) gas (and not the fraction
of the total mass, atomic plus molecular) that is dark.  It is also not the ratio
of the dark gas to the gas in the CO zone.

The gas in the cloud generally has large density fluctuations. In the \hone, this
is because the gas can exist in two distinct phases, cold ($T\sim 10^2$~K) and dense, and
warm ($T\sim 10^4$~K) and tenuous  (e.g., \citealp{wolfire03}), whereas the 
molecular gas is generally supersonically turbulent \citep{larson1981} and therefore has 
large density variations.
Let $n(\vecr)$ be the density of H nuclei at a point in the cloud, and
let $\bar n(r)$ be the locally volume-averaged density at that radius in the cloud.
Following \citet{larson1981}, we assume that $\bar{n}(r)\propto 1/r$.
This corresponds to a global column density that is independent of radius,
\beq
\bar{N}\equiv \frac{M(r)}{\muh\pi r^2}=2 r \bar{n}(r)=\mbox{const},
\label{eq:barN}
\eeq
where $\muh=2.34\times 10^{-24}$~g is the mass per H nucleus.
For numerical evaluation, we usually use a column density
$\bar{N} = 1.5\times 10^{22}$~cm\ee\  
from \citet{solomon1987}. The column $\bar{N}$ corresponds to an average 
visual extinction $\bar {A_V}$ through the cloud given by
\begin{equation}
\bar {A_V} = {{\bar N Z'} \over {1.9 \times 10^{21}\ {\rm cm^{-2}}}} \equiv 
\frac{\bar N}{N_0}= 5.26 \bar N_{22}Z',
\label{eq:av}
\end{equation}
where $N_0= 1.9\times 10^{21}Z'^{-1}$ cm$^{-2}$
is the radial column density corresponding to a visual extinction of unity
\citep{bohlin1978,rachford2002}, $\bar{N}_{22} \equiv \bar{N}/(10^{22}$~cm\ee),
 $Z'\equiv Z/Z_\odot$ is the metallicity relative to solar, and assuming
that the gas to dust ratio is constant within the cloud and
the dust opacity scales with the gas metallicity $Z'$.
The relation between the mass and
the locally volume-averaged density is 
\beq
\bar{n}(r)=  30.4\;\frac{\bar{N}_{22}^{3/2}}{{M_6(r)}^{1/2}} ~~\,\,{\rm cm^{-3}},
\label{eq:barnc}
\eeq
where 
$M_6(r)\equiv M(r)/(10^6\;M_\odot)$. 

The PDR code is run for a cloud of mass $M(\rco)$ 
on a grid of visual extinction, $A_V$, measured from 
the surface of the cloud at $R_{\rm tot}$ to a point $r$ inside
the cloud.  
We denote the corresponding column density as
\beq
N(r)\equiv - \int^r_{R_{\rm tot}}\bar n dr'
\label{eq:ns}
\eeq
 to distinguish it
from the globally averaged value in equation (\ref{eq:barN}).
Note that $N(r)$ is the column density {\it outside} of $r$,
whereas $\bar N$ is the average column density through the cloud and does
not depend on $r$ for our $1/r$ average density distribution.
The visual extinction $A_V$ corresponding to a column density $N$ is
given as in equation (\ref{eq:av}) with $\bar N$ replaced by $N$.
Note that
the column density, $N$ (or extinction, $A_V$)
between $\rco$ and $\rtot$ is not known {\it a priori} and 
is one of the main results of our calculation. In addition, for a 
$1/r$ mean density distribution, the column through the cloud center 
is infinite, but the total mass within the $\tau_{\rm CO}=1$ surface
is well defined and is one of the input parameters. In practice we run 
the PDR model to a depth of $A_V=10$
and adjust the inner radius of our grid at $A_V = 10$
so that the integrated CO
cloud mass, $M(\rco)$, is the input value. Our results do
not depend on the value of $A_V$ we choose for the inner radius since
there is so little mass or radius contained in the small central region where the
radial $A_V$ goes to infinity as $r$ goes to zero.

The results from the PDR code determine both
the extinction at the \hone\ -- \htwo\ interface, $A_V(\rht)$ and
the extinction at the surface of the CO cloud, $A_V(\rco)$.
The difference between these gives the extinction of the dark-gas layer,
\beq
\davdg\equiv A_V(\rco)-A_V(\rht).
\eeq
For a cloud with an $r^{-1}$ density profile, corresponding to $M(r)\propto r^2$,
the dark-gas mass fraction is
\begin{eqnarray}
f_{\rm DG}&=&1-\left(\frac{\rco}{\rht}\right)^2\, ,\\
&=& 1-\exp\left(-\frac{4N_0\davdg}{\bar N}\right),
\label{eq:fDGNo}\\
&=& 1-\exp\left(\frac{-0.76  \davdg}{Z'\bar{N}_{22}}\right)\\
\label{eq:fDGN}
&=& 1-\exp\left(\frac{-4.0  \davdg}{\bar{A}_V}\right).
\label{eq:fDGAV}
\end{eqnarray}
This result is generalized to other density profiles in Appendix \ref{appen:den}.
The dark-gas fraction thus depends upon only two quantities,
the radial column density of the dark-gas layer, $\Delta  N=N_0\davdg$, 
and the mean column density of the cloud, $\bar{N}$. When
expressed in terms of the extinction of the dark gas layer,
$\davdg$, a dependence on the metallicity, $Z'$,
also enters. Perhaps most simply, when expressed in terms of $\davdg$ and
$\bar A_V$, the fraction depends only on the ratio $\davdg / \bar A_V$.   This
is intuitive since (discussed later in detail), $\davdg$ is a measure of the dark
gas mass and $\bar A_V$ is a measure of the molecular mass.

\section{
Radiatively Heated Clumps at Constant Pressure}
\label{subsec:OpticallyThinConstantP}

As a first step, we consider the case of isobaric clumps with a thermal pressure
that is independent of position within the cloud. The clumps are
heated by an FUV/soft X-ray radiation field that is attenuated by
the gas between the clumps and the surface of the PDR, plus a column
of $10^{19}$~cm\ee.
(The clumps are also heated by the cosmic rays, but this heating is generally less
important in the dark-gas region.) 
The thermal pressure is $P_{\rm th}=x_t n_c kT$,
where $x_t$ is the sum over the 
abundances of all
species relative to hydrogen nuclei,
$x_t = \Sigma_i n_i/n$;
for atomic gas $x_t \simeq 1.1$, while for molecular gas 
$x_t\approx 0.6$. 
Note that the density derived from the isobaric PDR model is $n_c$, the density of the 
cold ($T \al 300$ K) gas component.
We assume that warm ($T\sim 8000$ K) \hone\ at the same thermal pressure
fills the remaining space,
but we assume that this has a negligible fraction of the mass.
As noted in \S \ref{sec:Models} above, the volume filling factor is given by 
$f_{\rm V}(r) = \bar{n}(r)/n_c(r)$, 
where $\bar{n}(r)$ is the locally volume-averaged H nucleus density 
(i.e., the density at $r$ averaged over the clumps since we are
neglecting the interclump medium)
and $n_c(r)$ is the 
H-nucleus density of the cold (clump) gas.

The model outputs include the local density in  the clumps, $n_c(r)$,
the volume filling factor
of the clumps, $f_{\rm V}$,
the fraction of atomic hydrogen,
$x_{{\rm HI}}= n_{{\rm HI}}/n_c$, and the fraction of molecular hydrogen, 
$x_{\rm H_2}= n_{{\rm H_2}}/n_c=\frac 12 (1-x_{{\rm HI}})$ 
as functions of extinction from the surface of the cloud,
$A_V$.
The total mass in each component is found by integrating the density
distributions
\beq
 M(\rht) = \int_0^{R_{\rm tot}} 2\muh\, 
   x_{\rm H_2}(r') n_c(r') f_{\rm V}(r') 4\pi r'^2\, 
    dr'\,\,\,\,\, ,
\label{eq:massH2}
\eeq
\beq
 M(\rco) = \int_0^{\rco} 2\muh\,  
    x_{\rm H_2}(r') n_c(r') f_{\rm V}(r') 4\pi r'^2\, 
    dr'\,\,\,\,\, ,
\label{eq:massCO}
\eeq
where $\rco$ is the radius of the $\tau_{\rm CO} = 1$ surface, 
and $\rtot$ is the outer radius of the entire cloud, which includes (from center outwards)
the CO region, the region with H$_2$ and C$^+$, and the outer atomic
envelope with atomic H and C$^+$ (see Fig.\ \ref{fig:avdiag} for 
an illustration of the various radii and optical depths).
Note that $\rco$ is an input
to our model while $R_{\rm tot}$ is an output; it is the outer radius that gives
 just enough shielding such that CO $J$=1-0 becomes optically 
thick at $\rco$.

We have run cases for a representative cloud of  mass 
$M(\rco) = 1\times 10^6$ $M_{\odot}$, with incident 
radiation fields $G_0'=\zeta_{\rm XR}'=10$, 
and two fixed values of the pressure, 
$P_{\rm th}/k = 10^4$ K ${\rm cm^{-3}}$ and $10^{5}$ K ${\rm cm^{-3}}$,
which covers the observed range of thermal pressure deduced from 
$^{12}$CO and $^{13}{\rm CO}$ observations of molecular clouds in 
the Galactic plane \citep{sanders93}. 
For $P_{\rm th}/k = 10^4$ K ${\rm cm^{-3}}$,
the atomic-molecular transition occurs at $A_V(\rht) = 0.54$, 
the transition to optically thick CO (at $\tau_{\rm CO}=1$) 
occurs at $A_V(\rco)=1.2$, and the dark-gas fraction
is found to be $f_{\rm DG} = 0.28$. 
The total molecular mass is related to $M(\rco)$ by
\beq
   M(\rht) = \frac{M(\rco)}{1-f_{\rm DG}}\; ,
\label{eq:mh2}
\eeq
with $M(\rht) = 1.4\, M(\rco)$ in this case.
For $P_{\rm th}/k = 10^5$ K ${\rm cm^{-3}}$,
we find $A_V(\rht) = 0.10$, 
$A_V(\rco)=0.86$, $f_{\rm DM} = 0.31$
and $M(\rht) = 1.4\, M(\rco)$. 
Although the transitions to ${\rm H_2}$ and CO gas
are shifted to lower column densities  
for the $P_{\rm th}/k = 10^5$ K ${\rm cm^{-3}}$
case, the dark-gas faction remains relatively 
unchanged. 

To investigate the sensitivity of this result to variations in the density 
distribution and FUV radiation fields, we have constructed
more realistic cloud models that include the effects of a
two-phase thermal pressure and the effects of turbulence.

\section{
Clumps in a Turbulent, Two-Phase Medium}
\label{subsec:OpticallyThinP2}

The illumination of the cloud by X-ray/EUV and FUV radiation
will heat and evaporate the clumps,
leading to a two-phase region 
in the outer regions of the molecular cloud
consisting of warm
$T\sim 8000$ K interclump gas and cooler $T \la 300$ K clumps. We assume 
the two-phase thermal pressure
 is the geometric mean of the minimum and maximum thermal pressures
allowed for a two-phase medium,
$\pthtp= (P_{\rm max}P_{\rm min})^{1/2}$,
and we again assume that the mass of the interclump gas is negligible.
Shocks or turbulence  will tend
to keep the thermal pressure above $P_{\rm min}$,
while condensation of WNM into CNM
will tend to keep the thermal pressure below $P_{\rm max}$.
The geometric mean
is roughly consistent with the mean thermal pressure in cold gas
found in  
numerical simulations of two-phase ISM disks with turbulence
\citep{piontek2007}.
For sufficiently strong external radiation fields, 
photoevaporation of the cool clumps 
near the outer boundary of the cloud
increases the density (and therefore
the thermal pressure) of the all-pervasive interclump medium, so that
the thermal pressure due to radiative heating exceeds the turbulent pressure there.  
However, the thermal pressure due to radiative heating drops
from the cloud surface inward due to absorption of the soft X-ray/EUV and
FUV radiation. 

For sufficiently weak radiation fields
such as exist in the interior regions of molecular clouds,
most of the gas is cold and the 
turbulent motions are very supersonic,
which generates a wide range of thermal pressures.
Both observational \citep{lombardi2006,ridge2006} and theoretical
\citep{vazquez1994,ostriker2001} 
studies suggest that turbulent clouds
have a log-normal density distribution that applies to
the dense component in the two-phase medium \citep{audit2010}.
The mass-weighted median density, $\langle n \rangle_{\rm med}$, 
in this turbulent medium is 
related to the volume-averaged density, $\bar{n}$, by 
\beq
  \langle n \rangle_{\rm med} = \bar{n} \exp(\mu)\,\, , 
\label{eq:nmed}
\eeq
where 
\beq
\mu \approx 0.5 \ln (1+0.25{\cal M}^2)\; ,
\label{eq:mu}
\eeq
\citep{padoan1997},
${\cal M}=\sqrt 3\sigma/c_s$ is the 3D Mach number, $\sigma$ is the 1D velocity dispersion, and $c_{\rm s}$ is 
the sound speed. The typical thermal pressure in a turbulent medium is
$\pthturb=x_t \langle n \rangle_{\rm med} kT$: half the mass is at a greater thermal
pressure than this and half at a lower one.
To treat the complex situation in which both radiative heating and turbulence determine
the pressure, we assume that the typical thermal pressure in the gas is
the larger of that due to radiative heating, $\pthtp$, and that due to
turbulence, $\pthturb$,
\beq
\pth=\max\left(\pthtp,\pthturb\right).
\eeq

To determine the Mach number, we note that the relation between
the linewidth and size of a cloud is given by the identity
\beq
\sigma(r)=0.52\left[\alpha_{\rm vir}
\left(\frac{\Sigma}{10^2 \;M_\odot\;\mbox{pc\ee}}\right) \left(\frac{r}{\rm pc}\right)\right]^{1/2}
~~~\mbox{km s\e},
\eeq
where $\alpha_{\rm vir}\equiv5\sigma^2r/GM$ is the virial parameter
and $\Sigma\equiv M/\pi r^2=\muh\bar{N}$ is the average mass surface density of the cloud
(see eq.\ 27 in \citealp{ostriker07}). 
\citet{heyer2009} emphasized the importance of the
dependence of the linewidth on column density and showed that the data on Galactic molecular
clouds are consistent with this scaling with $\alpha_{\rm vir}=1$.
By themselves, the GMCs in \citet{solomon1987}'s sample do not
show clear evidence for the $\bar N^{1/2}$ scaling; that becomes
evident only when higher resolution data are included. Nonetheless, 
normalizing the \citet{heyer2009} relation so that it agrees with
the linewidth-size relation of \citet{solomon1987} 
at the mean column density of the latter's sample ($\bar N=1.5\times 10^{22}$~cm\ee)
gives a linewidth-size relation
\beq
\sigma(r) = 0.72 \left(  \frac{\bar{N}_{22}}{1.5} \right )^{1/2}\left 
    ( \frac{r}{\rm pc}\right )^{1/2}
\,\,\, {\rm km\,\, s^{-1}}\, ,
\label{eq:vturb}
\eeq
We adopt this relation for the 1D velocity  dispersion here.
The corresponding 3D Mach number is ${\cal M}=\surd 3 \sigma(r)/c_s$,
where the isothermal sound speed is
$c_{\rm s}(r) = [x_tkT/(\muh)]^{1/2}$.

At each $A_V$ step we use a 
precalculated lookup table to find the appropriate two-phase thermal 
pressure $\pthtp=(P_{\rm max}P_{\rm min})^{1/2}$,  as a function of total column 
density $N$ from the cloud surface and 
molecular fraction  $f({\rm H_2}) = 2N_{\rm H_2}/ N$
(see Fig.\ \ref{fig:paveplt}). The pressure is calculated in a 
manner similar to \cite{wolfire03} except that here we include
the ${\rm H_2}$ self-shielding to depth $A_V$. The maximum, $\pmax$, and minimum,
$\pmin$, 
thermal pressures for a two-phase medium are found from phase diagrams
of thermal pressure versus density. 
In the regime where two-phase pressure
dominates, the gas temperature and local density $n_c(r)$  
are found 
self-consistently by iteration. In addition, however, we need
the mass-weighted
median density in the turbulent medium,
$\langle n \rangle_{\rm med}$ (eq.\ \ref{eq:nmed}), 
which is a function of the 
cloud temperature $T(r)$ [through $c_{\rm s}(T)$], and the 
cloud radius [through $\sigma(r)$, $\bar{n}(r)$, and $T(r)$].
The solution
requires two sets of iterations. First, we iterate between
$\langle n \rangle_{\rm med}$ and $T$ at each $A_V$ grid point to find
a self-consistent density. Second, we convert the $A_V$ grid to 
a radius grid and rerun the temperature, density iteration. 
 
We note that the ${\rm H_2}$ formation rate per unit volume is proportional to  
$ n_{\rm HI}n$, so one might expect 
that the mass-weighted mean density
$\langle n \rangle_{\rm M} = 
\langle n^2 \rangle_{\rm V}/\bar{n}=
\bar{n} \exp (2\mu)$  would be appropriate for 
${\rm H_2}$ formation and thermal balance rather than
the median density, $\langle n \rangle_{\rm med} = \bar{n} \exp (\mu)$;
here the subscript ``V'' refers to the volume average. 
However,
in highly supersonic gas,
only a very small fraction of the mass is above the 
mass-weighted mean density, so
the bulk of the chemistry occurs in the lower
density gas. 
When the transition from atomic to molecular gas is sharp,
then the conditions for the transition are determined by 
the lower density atomic gas, not by the molecular gas. 
We also note that numerical simulations of 
\cite{glover2007} suggest that ${\rm H_2}$ formation
 proceeds rapidly in a turbulent medium where the 
${\rm H_2}$ forms in high density gas and remains
molecular when transported to a lower density region.
However, using $\langle n \rangle_{\rm M}$ for the typical density
at which the formation occurs
does not take into account the fact that in very dense
regions the gas becomes fully molecular, so that the 
formation rate drops; thus use of the higher density  
would overestimate the formation rate.  Therefore, we
use  $\langle n \rangle_{\rm med}$ for the typical density
of a clump.

\subsection{Results for Clumps in a Turbulent Medium}
\label{subsubsec:Results}

We have applied our optically thin clump models 
to molecular cloud masses 
$M(\rco) = 1\times 10^{5}$ ${M_{\odot}}$, 
$3\times 10^{5}$ $M_{\odot}$, 
$1\times 10^{6}$ $M_{\odot}$, and $3\times 10^{6}$ $M_{\odot}$ for
incident radiation fields 
$G_0'=\zeta_{\rm XR}' = 3$, 10, and 30. The lower end of
the mass range is chosen to be the lowest cloud mass that could still
produce OB stars sufficient to illuminate, heat, and dissociate the
molecular cloud with FUV radiation. The upper end is about 1/2 the maximum
mass cloud in the Galaxy \citep{williams1997}.
The range in radiation
fields generously covers the expected elevated field from the sum of nearby stellar
associations. 
 
We show in Figure \ref{fig:Preslg} ({\em top}) the calculated thermal pressures 
for a representative cloud of mass $M(\rco)=1\times 10^{6}$ $M_{\odot}$
with incident radiation fields $G_0'=\zeta_{\rm XR}'=10$. 
In this case, the two-phase thermal pressure 
(i.e., the thermal pressure due to radiative heating)
dominates in the outer portion of the cloud, but drops
initially due to absorption of EUV/soft X-rays and then due to absorption
of FUV by dust. We note that the drop in two-phase pressure is greater
than expected from Table 3 in \cite{wolfire03}. They quote a decrease by 
a factor $\sim 1.8$ for $\pthtp$ (given as $P_{\rm ave}$ 
in that paper)
 between $N=10^{19}$ cm$^{-2}$ and $N=10^{20}$
cm$^{-2}$, whereas we find a drop by a factor of $\sim 3.7$. The difference is
due to the treatment of cosmic rays: here we do not scale the  
cosmic ray ionization
rate with $G_0'$ but hold it fixed at the local Galactic value. Thus the 
ionization falls off faster with depth into the cloud than in the 
\cite{wolfire03} models. 
The turbulent thermal pressure dominates for
$A_V\ga 0.001$. 
The transition to ${\rm H_2}$ (at $n_{{\rm H_2}}/n=0.25$)
occurs at $A_V(\rht) = 0.34$, 
the transition to optically thick CO (at $\tau_{\rm CO}=1$) 
occurs at $A_V(\rco)=1.0$, and the dark gas fraction is $f_{\rm DG} = 0.28$.

The calculated densities are shown in Figure \ref{fig:Preslg} 
({\em bottom}). 
We denote the clump density that would occur in two-phase equilibrium in
the absence of turbulence by $n_{\rm 2p}$. For the actual clump density, we
take
\beq
n_c=\max(n_{\rm 2p},\,\langle n\rangle_{\rm med}),
\label{eq:ncmax}
\eeq
where the turbulent mass-weighted median density, $\langle n \rangle_{\rm med}$,
is given by equation (\ref{eq:nmed}).
For the case in the figure, we have
$n_c=n_{\rm 2p}$ for $A_V \la 0.001$ and 
$n_c=\langle n \rangle_{\rm med}$
for $A_V \ga 0.001$. 
Both ${\rm H_2}$ and CO form
well within the turbulence-dominated region, a result  
that holds for all cloud masses and radiation fields considered 
here.\footnote{We note that the density in the atomic
portion of the cloud is close to the value of 
$230$ ${\rm cm^{-3}}$ given by the theory of
\cite{krumholz2009} for this case ($G_0'=10,\;Z=1$).
Their result is closely tied to the present work, since they adopt 
a density of $3n_{\rm min}$,
where the expression for the minimum two-phase density,
$n_{\rm min}$, is given in \cite{wolfire03}.}
Note that the volume filling factor $\fv \equiv \bar n/n_c \sim 0.1$ in 
the outer part of the cloud where the thermal pressure dominates
and two phases exist.  Here, 
WNM fills the rest of the volume with a density
of $\sim 10^{-2} n_c$.   This implies the mass of the interclump
medium is negligible ($\sim 0.1$) compared to the mass in clumps,
as we have assumed.  In the turbulence-dominated region, there is no
such thing as an interclump medium, but rather a distribution of densities
that fill the volume.  Here, most of the mass is at densities near $\langle n\rangle_{\rm med}$,
as assumed.

Figure \ref{fig:templg}  also shows
the temperature distribution ({\em top}) and 
the distribution 
of abundances for ${\rm C^+}$, ${\rm C^0}$, CO, and OH ({\em bottom}). 
The CO amounts to only $\sim 30$\% of the total carbon abundance when 
the $J$=1-0 line becomes optically thick at a CO column density of 
$N({\rm CO})\approx 2\times 10^{16}$ cm$^{-2}$.  Note that we do 
not include freeze out of ${\rm H_2O}$ on grains, which would be 
important at $A_V \ag 3$ \citep{hollenbach2009}, so our
OH chemistry becomes somewhat unreliable at these depths. However, this
happens after CO is formed and will not change our results.

For comparison with the $M(\rco) = 1\times 10^6$ $M_{\odot}$ case 
we also show in Figures \ref{fig:Preslglm} and \ref{fig:templglm}
the results for $M(\rco) = 1\times 10^5$ $M_{\odot}$. Figure \ref{fig:Preslglm}
shows the thermal pressures and densities and Figure \ref{fig:templglm} shows
the gas temperature and abundances.
The lower cloud mass results in higher $\bar n$ (eq.\ \ref{eq:barnc})
and thus higher clump density (eq.\ \ref{eq:nmed}) and higher
turbulent thermal pressure. The turbulent thermal pressure
dominates the two-phase thermal pressure at all $A_V$.
We find $A_V(\rht) = 0.23$, $A_V(\rco)=0.95$, and the dark gas 
fraction $f_{\rm DG} = 0.30$.

The optical depths $A_V(\rht)$ and 
$A_V(\rco)$ are presented in Figure
\ref{fig:Rout} ({\em top}) for cloud masses $M(\rco) = 1\times 10^{5}$ 
$M_{\odot}$ and
$M(\rco) = 3\times 10^{6}$ $M_{\odot}$ and for 
$G_0'=\zeta_{\rm XR}'=3$, 10,
and 30. The optical depths are measured from the cloud
surface inward. Figure \ref{fig:Rout} ({\em bottom}) shows
the cloud radii (measured from the cloud center) corresponding
to $A_V(R_{\rm H_2}$), $A_V(\rco)$ and the total
cloud radius.
First, we see that the optical depths to both the
${\rm H_2}$ and CO layers increase with increasing radiation fields.
(i.e., the transition layers are at greater column densities from the
cloud surface). In addition, higher cloud masses result in (slightly) greater 
optical depths to the transition layers. 
However, Figure \ref{fig:Rout} ({\em top}) shows 
$\davdg$ 
to be  nearly constant
at $\sim 0.6-0.8$ over our entire parameter range. 

The cloud radii are shown in Figure \ref{fig:Rout} ({\em bottom}).
The radius in CO, $\rco$ is fixed from observations 
by the cloud mass
and column density,
(eq.\ \ref{eq:barN}) and is independent of 
the incident radiation field. 
The next largest radius, $\rht$, is also found to be relatively 
constant with radiation field, reflecting the constancy
of $\davdg$ (Fig.\ \ref{fig:Rout}, {\em top}).  
The largest radius,
$R_{\rm tot}$, which encompasses the atomic surface as well as the
molecular interior, is found to increase slightly with 
increasing radiation 
field, a consequence of the higher column densities required 
to turn the gas molecular. We also show in Figure \ref{fig:Routrat} the
ratio of radii $R_{\rm tot}/\rco$ and $\rht/\rco$ for cloud 
masses $M(\rco) = 1\times 10^5$ $M_{\odot}$ and 
$M(\rco) = 3\times 10^6$ $M_{\odot}$.

In Figure \ref{fig:ratio} we present 
the calculated dark gas fraction $f_{\rm DG}$ (see eq.\ \ref{eq:fDG})
for all of the cloud masses and incident radiation fields. We find the
fraction is remarkably constant with both cloud mass 
and radiation field at a value of $f_{\rm DG}\approx 0.3$.  
As shown in equation (\ref{eq:fDGN}),  the constant fraction follows directly 
from the constant  optical depth between the ${\rm H_2}$ and CO layers,
and the constant $Z^\prime$ and $\bar N$ (or $\bar A_V$)  assumed in this case.

\subsection{Interpretation of Results} 
\label{sec:interp}

The trends in $A_V(\rht)$, $A_V(\rco)$, 
$\davdg$, and $f_{\rm DG}$ 
can be understood
from the formation/destruction processes for ${\rm H_2}$ and CO
(Appendices \ref{appen:h2form} and \ref{appen:coform}) and 
the relation between $f_{\rm DG}$ and 
$\davdg$ (eq. \ref{eq:fDGN}).
In  Appendix \ref{appen:h2form}, 
we provide an analytic treatment of $A_V(\rht)$
by balancing the formation of H$_2$ on grains with FUV 
photodissociation of H$_2$.
There we assume that the density is uniform, 
and since most of the mass of the cloud is in the clumps or, in the
turbulent gas, at density  $\langle n \rangle_{\rm med}$, we use
the density $n_c$.  Strictly speaking, this density varies somewhat from $R_{tot}$
to $R_{\rm H_2}$, but since the solution heavily depends on what happens
at $R_{\rm H_2}$, we use the value of $n_c$ there.
The dependence on metallicity, $Z'\equiv Z/Z_\odot$, 
FUV field and density $n_c$
is given by equation (\ref{eq:avh2}): 
\begin{equation}
A_V(\rht) \simeq 0.142 \ln \left[ 5.2 \times 10^3 Z'
   \left({G_0' \over {Z' n_c}}\right)^{1.75}+1\right].
\label{eq:avh2main}
\end{equation}
Here, $n_c$ is in units of cm$^{-3}$.
We find that the column density to the transition region is
a function of $G_0'/n_c$ consistent with previous investigations
\citep[e.g.,][]{sternberg1988, hollenbach1999, wolfire2008,krumholz2008,mckee2010}.  

In Appendix \ref{appen:coform}
we provide an analytic treatment of $A_V(\rco)$ 
by balancing the formation of CO by gas phase chemistry 
with FUV photodissociation of CO. The dependence
on $G_0'$, $n_c$, and $Z'$ is given by equation (\ref{eq:avco}):
\beq
A_V(\rco) \simeq 0.102 \ln
 \left[ 3.3\times 10^7 \left( G_0'\over Z'n_c \right )^2 + 1\right]\, ,
\label{eq:avcomain}
\eeq
where we have substituted the numerical values for the constants
into the equation and used $T~\sim 50$ K from Figure \ref{fig:templg}.
Taking the difference of equations (\ref{eq:avcomain}) and 
(\ref{eq:avh2main})
we find:
\beq
\davdg= 0.53 - 0.045\ln 
    \left ( G_0'\over n_c \right )  - 0.097\ln\, (Z')\, ,
\label{eq:deltaavmainz}
\eeq
which leads to
\beq
\davdg= 0.53 - 0.045\ln 
    \left ( G_0'\over n_c \right )  
\label{eq:deltaavmain}
\eeq
for $Z'=1$.
Here we have assumed 
$G_0'/n_c > 0.0075Z'^{0.43}$ cm$^3$, as is usually the  case,
so that one can ignore the factor of unity term inside the brackets of 
equations (\ref{eq:avh2main}) and (\ref{eq:avcomain}).
As noted in Appendices \ref{appen:h2form} and 
\ref{appen:coform}, the analytic solutions for 
$A_{V}({\rm H_2})$ and 
$A_{V}({\rm CO})$ (for $Z'$ = 1) agree with
the numerical solutions to within 5\% and 15\% respectively.  The 
analytic solution for $\davdg$ (for $Z'=1$)
agrees to within 25\% of the model results (Fig.\ \ref{fig:Rout}).

The principal conclusion of this analysis is that
$\davdg$
is almost constant (eq.\ \ref{eq:deltaavmain}),
consistent with our numerical results in Figure \ref{fig:Rout}.
Equation (\ref{eq:fDGN}) shows that the dark-gas fraction, $\fdark$, 
is a function of only $\davdg$, which is nearly constant,
and of $Z'\bar N$ or $\bar A_V$.
For conditions typical of the Galaxy---$\davdg = 0.6-0.8$ (Fig.~\ref{fig:Rout}),
$Z'=1$ and $\bar N=1.5\times 10^{22}$~cm\ee (or $\bar A_V \simeq 8$)---we 
find $f_{\rm DG} = 0.26-0.33$, in
good agreement with the numerical results portrayed in Figure \ref{fig:ratio}. 
We now discuss the reasons behind these results.

\subsubsection{Insensitivity to CO Cloud Mass, $M(\rco)$}

Why is $\fdark$ so insensitive to the cloud mass?
From equation (\ref{eq:nmed}), in the limit
of large Mach number, the median density goes as
$\langle n \rangle_{\rm med} \propto \bar{n}{\cal M} 
\propto \bar{n} \sigma$.
A higher cloud mass results in a lower value of
$\bar{n} \propto 1/M(\rco)^{1/2}$ 
(eq.\ \ref{eq:barnc} with $\bar N_{22}$ constant), while the
turbulent velocity scales as $M(\rco)^{1/4}$. 
Thus $\langle n \rangle_{\rm med}
\propto 1/M(\rco)^{1/4}$ varies only weakly
with cloud mass, 
with lower densities leading to deeper dissociation layers
as the cloud mass increases.  However, as seen in equation 
(\ref{eq:deltaavmain}),
small changes in $n_c$ lead to very small changes in 
$\davdg$, and thus to very small changes in $f_{\rm DG}$ (see
eq.\ \ref{eq:fDGAV}).

\subsubsection{Insensitivity to Ambient Radiation Field, $G_0'$}
The dark gas fraction is also quite insensitive to variation in 
the incident radiation field at constant cloud mass,
$M(\rco)$.
For constant $M(\rco)$, 
the CO radius, $\rco$, and the
distribution and value of $\bar{n}(r)$ 
are unchanged, although the total cloud radius increases at the higher
radiation fields, and thus $\bar{n}(r)$ drops to lower values in the 
outer edges of the cloud. Since
$\davdg$ is also relatively constant with changing $G_0'$,
the ${\rm H_2}$ radius, $R_{\rm H_2}$, is unchanged. Since the 
radii $R_{\rm H_2}$ and $\rco$ are unchanged and the density distribution,
$\bar{n}(r)$, is unchanged, we find constant masses and mass fractions.
Alternatively, and perhaps more simply, equation
(\ref{eq:deltaavmain}) shows that
$\davdg$ is very weakly dependent on $G_0'$, and therefore
$f_{\rm DG}$ is also weakly dependent as long as $\bar A_V$ remains fixed
(eq.\ \ref{eq:fDGAV}).
We note that the ratio of dark-gas mass to {\it total} mass 
(including the \ion{H}{1}) does decrease with $G_0'$ as we add more 
shielding \ion{H}{1} material to maintain constant 
$M(\rco)$.

We have also investigated the effect of lowering the incident 
FUV and EUV/soft X-ray fields to $G_0' = \zeta_{\rm XR}' = 1$ for 
the $M(\rco)=1\times 10^6$ $M_{\odot}$ model. We find the dark gas
fraction drops slightly from $f_{\rm DG} = 0.28$ to 0.22. The ${\rm H_2}$ 
self-shielding becomes quite strong and draws the ${\rm H_2}$ transition
close to the surface, resulting in $A_V({\rm H_2}) \sim 0.02$, 
while $A_V({\rm CO})$ 
is $\sim 0.5$. Nevertheless, the fitted functions
(eqs.\ \ref{eq:avh2main} and \ref{eq:avcomain}) 
behave quite well and reproduce the model results to within
3\% and 8\% respectively. 

Decreasing the incident field to
$G_0'=0.5$, appropriate for an opaque cloud embedded in the ISRF, 
further decreases $f_{\rm DG}$ to $0.18$. 
We can check that we are getting
reasonable results in the $G_0' = 0.5$ case
by comparing with observations of $N_{\rm CO}$ versus
$A_V$.  For CO dark clouds,
\cite{federman1990} shows CO column densities 
of $N_{\rm CO}=4\times 10^{16}$ ${\rm cm^{-2}}$
(twice our fiducial $\tau_{\rm CO} = 1$ surface) at total cloud columns 
of $A_V\sim 1$, and   thus $A_V\sim  0.5$ to the
CO emitting gas. This is reasonably consistent with our $G_0' = 0.5$
model that finds $A_V(\rco) = 0.4$.
For ${\rm H_2}$ we note that steady state PDR models of 
diffuse gas \citep{wolfire2008} 
find good fits to the observed molecular fractions 
deduced
from UV absorption line studies. The densities ($n \sim 30$ ${\rm cm}^{-3}$)
in diffuse gas are lower than considered here. However, applying our fitted
formula for $A_V(\rht)$ to $G_0' =0.5$, and $n = 30$ cm$^{-3}$, 
yields $A_V(\rht)\sim 0.23$. This is in good agreement 
with the transition
to high molecular fractions $f({\rm H_2}) > 0.1$ found in the diffuse ISM at
$A_V \sim 0.26$ \citep{gillmon2006}. 

\subsubsection{Dependence on Cloud Column Density, $\bar N$}
\label{subsec:varN}

Equation (\ref{eq:fDGN}) shows that the dark-gas fraction is sensitive to
the mean column density of the cloud.
\cite{heyer2009} have argued that for Galactic molecular clouds,
the  mean column density within 
$\rco$ is about half the value we adopted from \cite{solomon1987},
or $\bar{N}_{22} = 1.5/2 = 0.75$.
As \citet{heyer2009} point out, this value is quite approximate, since it depends on
an uncertain correction for non-LTE effects in the excitation of $^{13}$CO.
Bolatto has found the same values for the mean column density 
($\bar N_{22} = 0.75$) and turbulent velocity in the SMC as Heyer adopts 
for the Galactic clouds. 

To investigate the effects of variations in the mean column density, we have also run 
$\bar{N}_{22} = 0.75$ models, including the $\bar{N}^{1/2}$ scaling for
$\sigma(r)$ 
suggested by \citet{heyer2009}
(eq.\ \ref{eq:vturb}).  
Figure \ref{fig:fdglz2} ({\em bottom}) shows the low column density 
($\bar{N}_{22} = 0.75$) results for 
$M(\rco) = 1\times 10^6$ ${ M_\odot}$ and $Z'=0.5$, 1, and 1.9.
We find higher dark-gas fractions in the $\bar{N}_{22} = 0.75$   models
compared to the $\bar{N}_{22} = 1.5$ models;
in particular, for
$Z'=1$, $\bar{N}_{22} = 0.75$, and
$G_0'=10$, the dark-gas fraction $f_{\rm DG}\sim 0.5$, or 
$M(\rht)/M(\rco)\sim 2$. The change in $f_{\rm DG}$ is almost
entirely due to the change in mean column density; the change
in $\sigma(r)$ accounts for only a few percent of the increase. 
There is currently no
evidence for such high values of dark-gas mass in the local Galaxy
for the high mass clouds that we model here 
and we favor the higher, $\bar{N}_{22} = 1.5$, value. 
We note that \cite{grenier2005} finds higher dark mass fractions in local
clouds with very low masses [$M(\rco) < 3\times 10^4$ $M_{\odot}$]. 
However, these
clouds when observed with high spatial resolution CO observations reveal
quite small mean column densities,  
$\bar{N}_{22} \al 0.2$ \citep{mizuno2001,yamamoto2006}
where we expect the dark gas fraction to be higher than
$f_{\rm DG} \sim 0.3$.

\subsubsection{Dependence on Metallicity, $Z'$}
\label{subsec:varZ}

In Appendices \ref{appen:h2form} and 
\ref{appen:coform} we explicitly give the dependence
of the ${\rm H_2}$ and CO formation rates on the metallicity relative to solar, $Z'$. 
For ${\rm H_2}$, the  metallicity enters due to the conversion from
$A_V$ to column density, 
$ N$, and also in the rate of formation of 
${\rm H_2}$, which proceeds on grain surfaces. For CO, the
metallicity enters in the $A_V$ to $ N$ conversion and also in 
the production of OH, which proceeds both on grains and in  gas-phase 
chemical reactions that depend on the gas-phase abundances 
of oxygen and carbon. 

We ran models for our standard
cloud mass, $M(\rco) = 1\times 10^6$ ${M_\odot}$, incident
radiation fields $G_0' = 3$ ,10, and  30, and for metal abundances $Z' = 0.5$, 
1, and 1.9.  The $Z'=0.5$ case
is appropriate for molecular clouds in the 
Large Magellanic Clouds \citep[LMC;][]{dufour1984}; the
$Z'=1.9$ case is appropriate for clouds in the Galactic Molecular
Ring at a Galactocentric radius of $4.5$ kpc \citep{rathborne2009}, based on 
an exponential scale length for the metallicity in the Galaxy of $H_{R}^Z = 6.2$ kpc
\citep{wolfire03}.
Figure \ref{fig:fdglz2} ({\em top}) shows that
$f_{\rm DG}$ increases with lower metallicity, as predicted
by equation (\ref{eq:fDGN}). In general, the extinction $\davdg$ 
between $\rht$ and $\rco$ is very weakly dependent
on $Z'$ (see eq.\ \ref{eq:deltaavmainz}); however, the column density 
of the dark gas region
goes roughly as $Z'^{-1}$ so that the mass in the 
dark gas increases with lower metallicity. 

There is some indication that the dust abundance in low metallicity 
systems such as the LMC scales as $Z'^2$ rather than linearly with the metallicity
$Z'$ \cite[e.g.,][]{weingartner2001}. 
To test the effects of this scaling 
we ran a model with $Z'=0.5$ for gas phase metals and
and a dust abundance $Z'^2=0.25$. We find $f_{\rm DG}$ increases to $f_{\rm DG}\sim 0.60$,
a value intermediate between a $Z'=0.25$ and $Z'=0.5$ scaling for
both dust and gas phase metals (see Fig.\ \ref{fig:fdglz2}).
As expected a decrease in the dust abundance at fixed
$\bar N$ decreases $\bar A_V=5.26 Z' \bar N_{22}$ leading
to higher dark gas fractions; however, the increase is limited
due to  the metal dependence of $A_V(\rco)$ which leads to smaller
$A_V(\rco)$ (and thus smaller $\davdg$)
at $Z' = 0.5$ than at $Z'=0.25$.   

We caution against extrapolating our results for varying $Z'$ to very low
values of $Z'$.  For fixed $\bar N$ and fixed $G_0'$, lowering $Z'$ reduces
the size of the molecular interior until first, the entire cloud becomes 
optically thin
in the CO $J=1-0$ transition, and finally, the entire cloud becomes atomic H
and C$^+$.  We shall examine the dark-gas fraction for $Z'\ll 0.5$ clouds
in the universe in a subsequent paper.

\subsubsection{Dependence on Visual Extinction Through the Cloud, 
$\bar A_V$}

Equation (\ref{eq:fDGAV})
shows that $f_{\rm DG}$ depends only on $\davdg / \bar A_V$, 
where $\bar A_V\equiv 5.26 Z' \bar N_{22}$ is the mean visual extinction 
through the cloud.  As discussed
above, $\davdg$ is only weakly dependent on $Z'$, $\bar N_{22}$, $G_0$ and $n_c$.
Therefore, the
main parameter that controls the dark gas fraction is the visual extinction
through the cloud,
$\bar A_V$. Figure \ref{fig:avmean} clearly shows this dependence; it plots  
$f_{\rm DG}$ against $\bar A_V$ for
two different values of $Z'$= 0.5 and 1.9, holding $G_0'$ fixed at 10 and 
adjusting the gas
density (see eqs.\ \ref{eq:barnc}, \ref{eq:nmed}, \ref{eq:mu}, and \ref{eq:vturb}) 
as $\bar N$ changes.  Although, at 
fixed $\bar A_V$, 
 the $Z'=0.5$  case does have a slightly higher value
of $f_{\rm DG}$ than the $Z'=1.9$ case\footnote{This slight change is 
caused by the 
slight dependence of $\davdg$ on $Z'$ seen
in equation \ref{eq:deltaavmainz} and also because the different $Z'$ cases have different 
$\bar N$, which then means that
the gas densities change in the two cases.}, the main parameter controlling
 $f_{\rm DG}$ is $\bar A_V$.  If we fix $\bar N$, lowering $Z'$ lowers the 
$\bar A_V$ through the
 cloud, and this is the cause of the large change in $f_{\rm DG}$ seen in 
Figure \ref{fig:fdglz2}.  Essentially, lowering
 $\bar A_V$ raises the mass of the dark gas located in the surface 
region {\it relative to the mass of the
 interior CO gas, which is fixed in our model}.   For $\bar A_V=2$, there 
is considerable shielded H$_2$
 gas out to large radius around the fixed mass CO ``core".   For 
$\bar A_V=30$, the CO ``core" takes up 
 much of the cloud, and the dark gas is but a thin shell on the surface.
 
 We have fixed the mass in the CO region in our models because that is what
 is generally observed.  However, to better understand the dependence of
  $f_{\rm DG}$ on $\bar A_V$, it is easier to consider the alternate case 
of a cloud
  of fixed {\it total} mass (i.e., \ion{H}{1} + H$_2$ + CO mass) 
and variable $Z'$.
  As $Z'$ is lowered, the $\bar A_V$ through the cloud is lowered.   The radius
  of the CO region, $R_{\rm CO}$, shrinks because of the reduced shielding 
   in the
  cloud. The mass of the CO region drops, as $R_{\rm CO}^2$ since we 
  have assumed
  that $\bar n \propto r^{-1}$.   However, the $\davdg$ is relatively 
  constant through
  the dark gas shell around the CO, so the column through the shell 
  rises as $Z'^{-1}$.
  The mass in the H$_2$ dark gas, $M(R_{\rm H_2}) - M(\rco)$, changes 
  as $R_{\rm eff}^2$
  times the column in the shell, where $R_{\rm CO} < R_{\rm eff}< R_{\rm H_2}$.  
   Taking $R_{\rm eff}=R_{\rm CO}$
  to obtain the minimum mass in the dark gas, we see that the ratio of 
  dark gas mass
  to mass in the CO region goes as $Z'^{-1}$ in this case. Therefore,  
   $f_{\rm DG}$ increases
  as $\bar A_V$ decreases ($Z'$ decreases)  in this example.

\subsubsection{Variation with Cosmic-Rays}
\label{subsubsec:varcr}

The discussion of \htwo\ and CO in the Appendices is based
on the assumption that cosmic-ray ionization is not essential in 
determining the abundances of these molecules.
In order to test the effects of higher cosmic-ray ionization rates,
we start with our $G_0'=10$, $M(\rco) = 1\times 10^{6}$ $M_{\odot}$
model, and enhance the cosmic-ray ionization rate by a factor of 10 
from low columns to  $A_V=2$ and then drop it back to our 
standard value so
as not to overheat the cloud interior.
The enhanced rate tends to increase the production of
${\rm He^+}$ from cosmic-ray ionization and increase
the destruction of CO through reactions with ${\rm He^+}$.
The result is to drive the CO slightly deeper into the 
cloud, but to leave the H$_2$ surface at the same depth, thereby
slightly increasing the dark mass fraction. 
We find that $f_{\rm DG}$ increases from $\sim 0.28$ to $\sim 0.39$.

\subsection{Comparison With Observations}
\label{subsec:compobs}

First, we note that the dark gas fraction $f_{\rm DG} = 0.28$ for our standard
model is in good agreement with the observations of
\citep{grenier2005}, who found $f_{\rm DG} \approx 0.3$.
for the four local clouds in their sample that are more massive 
than $M(\rco) > 3 \times 10^4$ $M_{\odot}$.
\footnote{Here we refer to the mass within the CO emitting regions 
that \cite{grenier2005} has designated $M_{\rm H_2}$. For clouds more 
massive than $M(\rco) > 3\times 10^4$ $M_{\odot}$ they found
dark mass fractions, $f_{\rm DG}$,  approximately 0.1 for
Cepheus-Cassiopeia-Polaris, 0.1 for Orion, 0.6 for Aquila-Ophiuchus-Libra, and
0.3 for Taurus-Perseus-Triangulum. For lower mass clouds they found
0.5 for Chamaeleon, 0.8 for Aquila-Sagittarius, and 0.6 for Pegasus.
The dark gas fractions for high mass clouds found by \cite{abdo2010} 
are 0.30 for Cepheus, and 0.37
for Cassiopeia, and 0.36 for the low mass Polaris cloud.} 
We note that there is considerable scatter in their dark mass fraction
for these 4 clouds, which may indicate scatter in the average extinctions 
through the clouds, but that the
average value of the dark mass fraction is close to our theoretical prediction.
In addition, \cite{abdo2010} from observations of nearby resolved clouds
find an average value of $f_{\rm DG} \approx 0.34$ for their two massive
clouds.  
\cite{grenier2005} and \cite{abdo2010} find $f_{\rm DG}$ higher than 
0.3 in very low mass clouds
[$M(\rco) < 3\times 10^4$ $M_{\odot}$] where the mean extinction is observed 
to be much less
than our standard value and a higher $f_{\rm DG}$ is consistent with our prediction.
  
Next, we compare our numerical results with the observations of
\ion{H}{1} halos around molecular clouds in the Galaxy. 
\cite{wannier1983} observed \ion{H}{1} halos around 8 molecular
clouds and concluded they extend several parsecs beyond the CO
and have a thickness of a few parsecs. In addition, the halos are ``warm,'' 
having temperatures of at least $50-200$ K in order to be seen in emission
over the background.  We note that the use of ``warm'' in their title has been 
interpreted by others to mean gas at temperatures of $\sim 8000$ K, 
but in fact they
cite temperatures of order a few 100 K. 

\cite{andersson1991} and
\cite{andersson1993} 
included additional data and analysis and 
report \ion{H}{1} integrated intensities between 700 K km ${\rm s}^{-1}$ 
and 4300 K km ${\rm s}^{-1}$ and characteristic depths of about 4.7 pc.
Based on model fits, they estimate \ion{H}{1} densities 
$n_{\rm HI} \sim 25-125$ cm$^{-3}$ and temperatures 
$T\sim 50-200$ K. We note that they suggested that either the formation rate
of ${\rm H_2}$ was lower than the standard rate or the FUV field was
about 10 times the interstellar value in order to simultaneously match
the CO and \ion{H}{1} observations. At that time they favored the lower
${\rm H_2}$ formation rate since an increased FUV field would produce too
little OH compared to observation. However subsequent, work by
 \cite{hollenbach2009}
showed that OH is formed at greater abundances than previously thought when grain
surface chemistry is included. Thus, their suggestion that the FUV field
is enhanced near GMCs is consistent with a similar result by Wolfire, 
Hollenbach, \& McKee
(2010, in preparation) in modeling the average FUV field illuminating a 
star-forming molecular cloud.

For 
cloud masses in the range $M(\rco) = (1-10)\times 10^5$ ${M}_{\odot}$ 
and for $G_0'=10-30$,  
we find \ion{H}{1} 
integrated intensities
from 965 K km ${\rm s^{-1}}$ to 4100 K km ${\rm s^{-1}}$ and a range
in \ion{H}{1} halo
 thickness from 1 pc to 10 pc. These are in 
good agreement with the observations. At $A_V(\rht)$,
our temperature range 
$T=70-80$ K and density range $\bar n = 45-147$ cm$^{-3}$
agree with that of \cite{andersson1993}.
Therefore, we 
consider the model to be in very good agreement with these observations. 

We note that  CO-to-${\rm H_2}$ conversion factors
have been inferred
from gamma-ray and  far-infrared
observations, and also from virial mass estimates.
The gamma-ray \citep{strong1996} and far-infrared 
\citep{dame2001} estimates
include the ``dark gas''  and are
$X=1.9\times 10^{20}$ ${\rm cm^{-2}}$ $({\rm K\ km\ s^{-1}})^{-1}$
and $X=1.8\times 10^{20}$ $({\rm K\ km\ s^{-1}})^{-1}$ 
from gamma-ray observations 
 and far-infrared observations respectively. However, these values 
are Galactic
averages and observations of nearby resolved clouds show a
variable $X$ increasing towards the outer Galaxy \citep{abdo2010}. 
The Abdo et al.\  observations separate the ``dark gas'' and CO components
and generally show lower $X$ ratios 
[$X\sim 0.87\pm 0.05\times 10^{20}$ ${\rm cm^{-2}}$ $({\rm K\ km\ s^{-1}})^{-1}$].  
Depending on whether the dark gas is mixed with other components 
the $X$ value could be lower if mixed with \ion{H}{1} or raised if mixed
with CO.
The \cite{solomon1987} conversion factor was based on a virial cloud estimate
and for an idealized cloud would not include the ``dark gas''.
\cite{ostriker07} re-evaluated their conversion,
accounting for an 8.5 kpc distance to
the Galactic center and including the He mass, and found a revised 
Solomon et al.\ value of
$X=1.9\times 10^{20}$ ${\rm cm^{-2}}$ $({\rm K\ km\ s^{-1}})^{-1}$. In general, the 
$X$ ratios derived from virial mass estimates exceed those based on 
gamma-ray observations
from resolved clouds by factors of 1.5-3.0 \citep{abdo2010}.
However, turbulent mixing of the bright CO and dark gas could easily cause the virial
estimate to include some of the dark gas.  Therefore, one should note
that CO to H$_2$ conversion factors are sometimes normalized to implicitly include
the dark mass, even though it lies outside the CO gas.

\subsection{Justification of the Optically Thin Approximation}
\label{sec:justif}

One of the important approximations in our work is that the density
fluctuations in the molecular cloud---the clumps---are optically thin,
or $A_V < 1$ through each clump.
This approximation can now be justified in light of our results: the
fact that the dark-gas fraction depends primarily on $\bar A_V$,
the mean visual extinction through the entire cloud, but only weakly on the
density and radiation field, implies that it is insensitive to the
distribution of matter within the cloud. Consider an extreme example
of a cloud with very opaque clumps, namely a cloud with clumps so opaque that
their mean
extinction is equal to that of the entire cloud, $\bar A_V\sim 8$. 
We assume that the cloud with opaque clumps has the
same mean $\bar A_V$, and clump density, $n_c$,
as the cloud with transparent clumps.
For simplicity we assume that there is no radial variation in the properties
of the clumps. Let $a$
be the clump radius. Then the number of clumps, $\caln$,
is determined by the condition that the clumps occupy a fraction
$\fv$ of the volume: 
\beq
\caln a^3=\fv \rtot^3.
\eeq
The condition that the clumps have the same $\bar A_V$  as the cloud as
a whole gives 
\beq
n_c a=\bar n \rtot=\fv n_c\rtot,
\eeq
so that $a=\fv \rtot$. When viewed from the outside, the fraction of the cloud
that is covered by clumps (the projected covering factor) is then
\beq
C=\frac{\caln \pi a^2}{\pi\rtot^2}=1,
\eeq
as expected. From the perspective of a clump inside the cloud, this means
that about half the sky is covered by clumps, since, on average, the path length from
a clump to the surface of the cloud is half the path length through the entire cloud. The average
radiation field incident on a clump is therefore reduced by about a factor 2
compared to the radiation field incident on the cloud as a whole. 

We can now estimate the dark-gas fraction of a cloud of these opaque clumps.
Each clump will have a dark-gas fraction that is almost the same as that of
the entire cloud, modified only by the reduction in $G_0'$ by about a factor of 2.
Since the extinction of the dark gas layer depends only
weakly on the intensity of the radiation field, $\davdg\propto 0.045\ln G_0'$,
this modification is very slight for typical values of $\bar A_V\sim 8$
(see eq.\ \ref{eq:fDGAV}). This argument does not prove that clumps with
$\bar A_V$ intermediate between the large value $\sim 8$ assumed here and
the small values $\bar A_V < 1$ assumed in the rest of this work will also have the
same dark-gas fraction. However, because this range $1 < \bar A_V < 8$ is small
it strongly suggests that this is the case. We
conclude that inclusion of finite optical depth of clumps in a cloud is
unlikely to significantly alter our conclusions.

\subsection{Justification of Assumption of Steady State Chemistry}
\label{sec:justifchem}

Our models solve for the steady state abundances of the atomic
and molecular species. In this subsection we examine the assumption
of steady state chemistry in the turbulent molecular cloud surfaces 
by comparing the chemical and dynamical times at the cloud depth 
$A(\rht)$. The chemical timescale, $t_{\rm chem}$,  is the time 
for atomic gas to become completely molecular 
($n_{\rm H_2} = 0.5 \langle n\rangle_{\rm med}$), thus
$t_{\rm chem}= 0.5/({\cal R} \langle n\rangle_{\rm med})$ where
${\cal R}=3\times 10^{-17}Z'$ cm$^3$ s$^{-1}$
is the rate coefficient for ${\rm H_2}$ formation on 
dust grains (see Appendix \ref{appen:h2form}). To compute the dynamical 
timescale, $t_{\rm dyn}$, we require the characteristic
distance for the turbulence to bring molecular gas from the 
interior to the surface, and to bring
atomic gas from the surface to the interior, 
where the interior is $A_V$ greater than $A_V(\rht)$. 
This characteristic distance is 
$d_{\rm dyn} = 1.9\times 10^{21} A_V(\rht)/\bar n$. 
The dynamical timescale is this distance divided by the
turbulent velocity for lengthscale $d_{\rm dyn}$,
$t_{\rm dyn} = d_{\rm dyn}/\sigma(d_{\rm dyn})$ where
$\sigma$ is the 1D turbulent velocity dispersion (eq.\ \ref{eq:vturb}).

For a cloud mass $M(\rco)=10^5$ $M_\odot$, and $Z'=1$, 
we find $\bar n\approx 140$
${\rm cm^{-3}}$, and $\langle n\rangle \approx 600$ ${\rm cm^{-3}}$ at
$A_V(\rht) = 0.22$ and thus $t_{\rm chem} \sim 2.8\times 10^{13}$ s, 
$d_{\rm dyn}\sim 1.0 $ pc, and $t_{\rm dyn} \sim 4.3\times 10^{13}$ s. 
Comparing chemical and dynamical timescales we find 
$t_{\rm chem}/t_{\rm dyn}=0.65$. Similarly, for $M(\rco) = 10^6$ $M_\odot$
we find $t_{\rm chem}/t_{\rm dyn} = 0.48$. 
For $t_{\rm chem}/t_{\rm dyn} \al 1$, a steady state approximation is
valid since gas has time to reach steady state before 
significant turbulent mixing. We find ratios less than one, but
only marginally so, and thus we expect some affects from turbulence.
We note however that turbulence brings atomic gas inward 
past $A_V(\rht)$ into the molecular gas, and at the same time
it brings molecular gas to the surface. Thus, the first order effect
is to spread out the rise of $x_{\rm H_2}$, but not to move $A_V(\rht)$.
In addition, we have demonstrated in subsection (\ref{subsec:compobs}) that
the steady state model is in good agreement with observations, and thus
the effects of turbulence are modest.

\section{Discussion and Summary}
\label{sec:Discussion}

\subsection{Model Assumptions}

We have constructed models of molecular clouds to investigate the fraction of
gas that is mainly ${\rm H_2}$  and contains little CO. These conditions
exists where the CO
is photodissociated into C and C$^+$ but the gas is molecular ${\rm H_2}$ due to either
${\rm H_2}$ self-shielding or dust shielding. Such conditions can exist
either on the surfaces of molecular clouds or the surfaces of clumps contained
within such clouds. Observations indicate that the mass in this ``dark gas''
can be as high as 30\% of the total molecular mass in the
local Galaxy. Previous theoretical
plane-parallel models  of individual PDRs have indicated that 
such a layer should exist 
\citep[e.g.,][]{vandishoeck1988},
but here we construct models directed towards molecular clouds 
as a whole while including observational constraints on 
cloud mass, radius, average density, and line width and theoretical
considerations of the likely strengths of the UV fields impinging on GMCs. 

We assume that the surface of each cloud is isotropically illuminated over $2\pi$ 
steradians by 
a soft X-ray/EUV field and an FUV radiation field. We use the 
standard cosmic-ray ionization rate of $1.8\times 10^{-17}$ ${\rm s^{-1}}$
per hydrogen nucleus everywhere in the cloud for all cases but one test  case. 
There is evidence from observations of ${\rm H_3^+}$
that the cosmic-ray ionization rate is a factor of 10 higher in some portions of the diffuse ISM
\citep{indriolo2007},
but there is no indication that such rates apply in molecular
cloud interiors \citep{mccall1999}. In preliminary work of M.\ Wolfire et al.\ 
(2010, in preparation) we
find that the average FUV field on clouds is $\sim 20$ times the
local Galactic interstellar field. This elevated field arises from
the distribution of OB associations around the cloud.  

The distribution of temperature, density, and abundances within the
\ion{H}{1}, ${\rm H_2}$, and CO layers are calculated using the PDR code
of \cite{kaufman2006}. In constructing model clouds from the PDR output,
we impose the constraint from \cite{solomon1987} 
that the typical column 
density through the cloud is $\bar{N}_{22}= 1.5$, independent of cloud mass
and independent of radius inside a given cloud (since $\bar n \propto r^{-1}$).
The locally averaged 
density $\bar{n}$  and radius of the CO cloud, $\rco$, 
as functions of the CO-cloud mass, $M(\rco)$
follow from equations
(\ref{eq:barN}) and (\ref{eq:barnc}).
In our notation, $M(\rco)$ is the molecular mass contained within 
the $\tau_{\rm CO}=1$ surface (of the CO $J=1-0$ transition)
including the mass of H$_2$ and He, and $\rco$
is the radius of this CO photosphere. 
We impose a volume-averaged density distribution $\bar{n}(r)$ 
that behaves as $\bar{n}\propto 1/r$ throughout all layers of the cloud. 
Note that in regions where the thermal pressure $P_{\rm th}$
lies between $\pmin$ and $\pmax$ both warm ($T\ag 7000$ K) 
and cold ($T\al 500$ K) gas solutions are 
possible. In this regime we use the cold solution from our model results
with $n_c$ the density of the cold clumps
and $\bar{n}$ the average over cold and warm gas components.  The warm gas
fills the volume, but contains little of the mass.

We test two extreme limits for clumps within
the cloud, one in which all clumps are optically thin to the incident 
FUV field, and one in which all clumps are optically thick to the
incident FUV field. In the optically thin approximation,
the PDR model output as a function of $A_V$ gives directly the distribution
in $A_V$ throughout the cloud. 
In the optically thin limit
we test two different
 models for the cloud density distribution. First we
use constant thermal pressure models and second we include 
two sources of thermal pressure, radiative heating (which
by itself would lead to a two-phase equilibrium) and supersonic turbulence. 
The distribution
of two-phase thermal pressure is calculated as in \cite{wolfire03} and 
stored in a look-up table as a function of total column density and
molecular fraction (see Fig.\ \ref{fig:paveplt}). 
 This pressure drops as one moves into the cloud due to the absorption of the
 radiation responsible for heating the gas.
Supersonic turbulence is characterized by a mass-weighted median density
$\langle n \rangle_{\rm med}=\bar{n}\exp(\mu)$, 
with $\mu = 0.5\ln (1+0.25{\cal M}^2)$ \citep{padoan1997}.
When the two-phase pressure drops below 
$P_{\rm th}^{\rm turb}\equiv x_t\langle n \rangle_{\rm med} k T$, 
where $x_t$ is the sum over the abundances of all species relative 
to hydrogen nuclei, we assume that the turbulence maintains the gas at 
a thermal pressure $P_{\rm th}^{\rm turb}$.
The sound speed that enters in the Mach number is calculated from the
PDR model output while the turbulent velocity is given by the linewidth-size relation
(eq.\ \ref{eq:vturb}).

In the limit of very optically thick clumps (with optical depths at least
as large as the average optical depth through the cloud), we find that
the average radiation field on a clump ranges between about $G_0'/2$ to $G_0'$.
In this limit, the dark mass fraction of a clump is the same as the dark mass fraction
of the entire cloud.  Since the dark gas mass fraction of a spherical clump or
a spherical cloud is insensitive to the incident FUV field 
(see eqs.\ \ref{eq:fDGN} and \ref{eq:deltaavmainz}),
the
dark mass fraction of the cloud  does not change significantly compared with
the case of a cloud made up of optically thin clumps.

We have tested the steady state assumption for the chemistry by
comparing the time to form molecular hydrogen, $t_{\rm chem}$, with
the dynamical time, $t_{\rm dyn}$, for turbulence to bring molecular gas from
the interior to the surface and to bring atomic gas to the interior. 
We find $t_{\rm chem}/t_{\rm dyn} \sim 0.7$ for a cloud mass
of $M(\rco) = 1\times 10^5$ $M_{\odot}$
and $t_{\rm chem}/t_{\rm dyn} \sim 0.5$ for a cloud mass of 
$M(\rco)= 1 \times 10^6$ $M_{\odot}$. Thus we expect modest affects
due to turbulence; mainly to spread out the transition from atomic
to molecular hydrogen but not to move $A_V(\rht)$. We also note
that steady state models agree well with observations.  

\subsection{Dark-Gas Fraction}

For our standard cloud mass, $M(\rco) = 1\times 10^6$ $M_\odot$, 
and incident radiation field, $G_0' = \zeta_{\rm XR}' = 10$, we find
dark-gas mass fractions of $f_{\rm DG} = 0.28$ and $f_{\rm DG} = 0.31$ 
for constant thermal pressures of $P_{\rm th} = 10^5$ K ${\rm cm^{-3}}$ and 
$P_{\rm th} = 10^6$ K ${\rm cm^{-3}}$, respectively. These correspond
to a total molecular mass $M(\rht) \approx 1.4 M(\rco)$.
For models that include both thermal and turbulent
pressures, we find essentially the same results as for the cases with constant
thermal pressure. The variation in $f_{\rm DG}$ ranges from 0.25 
to 0.33 over a range in $G_0'$ from 3 to 30 and a range
in cloud mass from $10^5$ $M_\odot$ to $3\times 10^6$ $M_\odot$
(Fig.\ \ref{fig:ratio}). Figure \ref{fig:Preslg} shows the distribution in 
thermal pressures and densities and Figure \ref{fig:templg} shows the 
distribution in temperatures and chemical abundances for the 
standard model. 
 
The constant value of $f_{\rm DG}$ for fixed $\bar A_V$  can be understood 
from 
the analytic solutions for $A_V(\rht)$ and $A_V(\rco)$ in
Appendices \ref{appen:h2form} and \ref{appen:coform} 
and the expression for $f_{\rm DG}$ in equation (\ref{eq:fDGAV}).
In the limit of $G_0'/n > 0.0075 Z'^{0.43}$ ${\rm cm^3}$, both $A_V(\rht)$ and
$A_V(\rco)$ increase as $\ln(G_0'/n)$. 
Thus, the optical depth through
the ${\rm H_2}$ layer $\davdg$ 
is a weak function of $G_0'/n$ and $Z'$ (eq.\  \ref{eq:deltaavmainz}) 
and is nearly constant over our parameter space. 
Furthermore, we
find that $f_{\rm DG}$ is a function of only $\davdg$ and the mean extinction
through the cloud $\bar A_V \equiv 5.26 Z' \bar N_{22}$ (Eq. \ref{eq:fDGAV});
thus, for a given $\bar A_V$, the dark-gas fraction is constant. 
However,  $f_{\rm DG}$ increases significantly if $\bar A_V$ decreases.

Our numerical results compare well with observations of the local 
Galactic dark-gas fraction, which \cite{grenier2005}
find to be $f_{\rm DG} \sim 0.3$,
 when averaged over their four most massive
clouds with masses between $3 \times 10^4 - 3\times 10^5$ M$_\odot$. 
Lower mass clouds observed by \cite{grenier2005} are observed to have
low $\bar A_V$ and thus high dark-gas fractions consistent
with our prediction. 
Our \ion{H}{1}
integrated intensities  of 965 K km ${\rm s^{-1}}$ to
4100  ${\rm s^{-1}}$ , \ion{H}{1} cloud-halo thickness of 1 pc to 10 pc,
average cloud densities of $\bar n \sim 45-150$ cm$^{-3}$, and average
\ion{H}{1}  temperatures of $\sim 70-80$ K
are in good agreement with the observations of \ion{H}{1}
cloud halos observed by \cite{wannier1983}, \cite{andersson1991}, and
\cite{andersson1993}.

We have carried out several additional tests to assess the dependence of
our results on the cosmic-ray ionization rate, 
the clump optical depth, the mean 
cloud column density,  the metallicity, and the mean visual extinction through the cloud. 
We ran our standard model with the cosmic-ray ionization rate a factor
of 10 higher at $A_V< 2$, as suggested by ${\rm H_3^+}$ observations in 
some regions of the
diffuse ISM, and found that this enhanced cosmic-ray 
rate only slightly increases
the dark-gas fraction. 
In the limit of very optically thick clumps we find no significant
change in $f_{\rm DG}$.

\cite{heyer2009} has suggested that the mean column density (and therefore the
mean visual extinction $\bar A_V$) in Galactic CO clouds
is about half the value found by \cite{solomon1987}, i.e.,  $\bar{N}_{22} \simeq 0.75$ or
$\bar A_V \simeq 4$. 
The results of changing the mean column density, the metallicity, and the mean visual
extinction through the cloud are 
illustrated in Figures \ref{fig:fdglz2} and \ref{fig:avmean}. 
We find that
the dark-gas fraction increases with lower extinction
since the dark gas occupies a larger fraction of the cloud 
(see eq.\ \ref{eq:fDGN}).
There is also a weak dependence of $\davdg$ on the mean column density
that tends to slightly mitigate the dominant effect. 
Lower mean columns lead to lower mean densities (eq.\ \ref{eq:barnc})
and lower $n_c$ (eqs.\ \ref{eq:nmed} and \ref{eq:ncmax}), and
thus lower $\davdg$ (eq.\ \ref{eq:deltaavmainz}).
We examine metallicities appropriate
for the LMC ($Z' = 0.5$), for the local Galaxy ($Z'=1$), and for the molecular
ring at $R=4.5$ kpc ($Z'=1.9$). 
In general, $f_{\rm DG}$ increases as the
metallicity drops for fixed columns because the mean extinction through the cloud
decreases, which raises the ratio of the surface dark gas to the interior CO gas (again, see eq.\ \ref{eq:fDGN}).
There is also a weak dependence of $\davdg$ on $Z'$ that also
slightly increases $\davdg$ (or $f_{\rm DG}$)  with decreasing $Z'$ 
(eq.\ \ref{eq:deltaavmainz}) even if the column is changed so that $\bar A_V$ remains fixed.
We note that in the case of varying $\bar{N}$ and $Z'$, but at
constant $\bar{N}Z'$ or $\bar A_V$ (see Figure \ref{fig:avmean}), the change in $f_{\rm DG}$
is entirely due to the weak dependencies of $\davdg$ on $\bar{N}$
and $Z'$ as noted above and shown in equation (\ref{eq:deltaavmainz}).
We also examine the case for dust scaling as $Z'^2$ ($=0.25$) while gas phase 
metals scale as $Z'$ ($=0.5$). We find $f_{\rm DG}$ is larger  
than when both metals and gas scale together. 

In Appendices \ref{appen:h2form} and \ref{appen:coform} we derive analytic
solutions for $A_V(\rht)$ and $A_V(\rco)$ as functions of density 
$n$, FUV field $G_0'$, and metallicity $Z'$. For the case of ${\rm H_2}$
we find an
expression for the abundance of ${\rm H_2}$ by balancing formation
and destruction processes and then solving for the position where  
$n_{{\rm H_2}} = 0.25 n$. Similar studies have been carried out by, for example, 
\citet{sternberg1988,mckee2010}. We find our fits 
are good to $\pm 5$\% for all models presented in this paper. The fits
to CO are found by integrating the expression for the abundance
of CO to a column density of $N({\rm CO})= 2\times 10^{16}$ ${\rm cm^{-2}}$,
where  $\tau_{\rm CO} = 1$ (the optical depth of the CO J= 1-0 transition). 
The fits are generally good to within
$\pm 15$\% except for the lowest $Z'$ and $G_0'$ model, where the 
fit is good to $\pm 25$\%. 
We have also derived analytic expressions for $\davdg$ 
and $f_{\rm DG}$ in the main text (eqs. \ref{eq:deltaavmainz} and \ref{eq:fDGN},
respectively).

The overall result of this paper is a theoretical derivation 
that $f_{\rm DG}\simeq 0.3 \pm 0.08$ for
GMCs with $\bar A_V \simeq 8$ in our Galaxy.  Therefore, a significant fraction of the molecular
gas in our Galaxy lies outside the CO gas.  As discussed in \S 2.4.2, some
calibrations of the CO line intensities to H$_2$ mass take into account this
``dark" H$_2$ gas.  However, it is important to be aware of this component
since it contributes significantly to the gamma ray, infrared/submillimeter continuum,
and [CII] 158 $\mu$m emission from clouds in galaxies.  
Its contribution to the star formation in a galaxy is as yet undetermined.  
The importance of this component increases as
the metallicity decreases, such as in the outer regions of galaxies or in
low metallicity galaxies.

Although the gas in the C$^+$/${\rm H_2}$ layer is termed
``dark gas,'' it emits [\ion{C}{2}] 158 $\mu$m line emission
and the dust in the dark layer
emits infrared continuum.
In a subsequent paper we will estimate the emission from the ``dark gas''. 
In addition, a future paper will model in detail the lower metallicity 
clouds found, for example, in the SMC and early universe. 

\acknowledgements

Partial support for M.G.W. and D.J.H. was provided by a NASA Long Term Space 
Astrophysics  Grant NNG05G64G. Partial support for D.J.H was also provided by
NASA's Stratospheric Terahertz Observatory project, grant NNX08AG39G.  Support for CFM was
provided in part by the National Science Foundation
through grant AST-0908553. We thank Alberto Bolatto and 
Eve Ostriker for useful discussions. We also thank an anonymous
referee for a careful reading of this manuscript and for several
very helpful comments.


\appendix

\section{Dark-Gas Mass Fraction}
\label{appen:den}

Here we determine the dark-gas mass fraction, 
\beq
\fdark\equiv\frac{M(\rht)-M(\rco)}{M(\rht)}\; ,
\label{eq:fdgappa1}
\eeq
for clouds with a power-law gradient for the mean density,
\beq
\bar n=\bar n(\rht)\left(\frac{\rht}{r}\right)^\krho,
\eeq
for $\krho<3$ (the reason for normalizing the density at
$\rht$ will become apparent below). GMCs typically have $\krho\simeq 1$ \citep{larson1981},
which is the case considered in the text. For $\krho<3$, the mass inside
a radius $r$ is
\beq
M(r)=\frac{4\pi \bar n(\rht)\muh \rht^\krho}{3-\krho}\; r^{3-\krho},
\label{eq:mass}
\eeq
and the mean column density inside $r$ is
\beq
\bar N(\rht)\equiv \frac{M(\rht)}{\muh \pi \rht^2}=\frac{4\rht\bar n(\rht)}{3-\krho}\; ,
\label{eq:barNapp}
\eeq
where $\muh$ is the mean mass per H nucleus.
For $\krho\neq 1$, the column density of the dark-gas layer is
\begin{eqnarray}
\dnsdg&=&\bar n(\rht)\int_{\rco}^{\rht}\left(\frac{\rht}{r'}\right)^\krho\; dr',\\
&=&\bar N(\rht)\left(\frac{3-\krho}{\krho-1}\right)\left[\left(\frac{\rht}{\rco}\right)^{\krho-1} -1\right],
\label{eq:nsapp}
\end{eqnarray}
where we used equation (\ref{eq:barNapp}) in the second step.
The mass fraction of the dark gas (eq.\ \ref{eq:fdgappa1}) becomes
\beq
\fdark=1-\left(\frac{\rco}{\rht}\right)^{3-\krho}
\eeq
with the aid of equation (\ref{eq:mass}). Equation (\ref{eq:nsapp}) then implies
\begin{eqnarray}
\fdark&=&1-\left[ 1+ 4\left(\frac{\krho-1}{3-\krho}\right)\frac{\dnsdg}{\bar N(\rht)}\right]^{-(3-\krho)/(\krho-1)}\, ,\\
&=&1-\left[ 1+ \left(\frac{\krho-1}{3-\krho}\right)\frac{0.76\davdg}{Z'\bar N_{22}(\rht)}\right]^{-(3-\krho)/(\krho-1)}\; ,
\label{eq:fdgapp}
\end{eqnarray}
where we used equation (\ref{eq:av}) in the second step.
Taking the limit as $\krho\rightarrow 1$ gives equation (\ref{eq:fDGNo}) in the text, since
$\dnsdg=N_0\davdg$ and $\bar N$ is constant, so the argument $\rht$ is not needed.

In order to see how the dark-gas fraction depends on the density gradient, we
rewrite equation (\ref{eq:fdgapp}) as
\beq
\fdark=1-\left(1+\frac{x}{\alpha}\right)^{-\alpha}\; ,
\label{eq:appenafdg}
\eeq
with
\begin{eqnarray}
x&\equiv &\frac{4\dnsdg}{\bar N(\rht)}=\frac{0.76\davdg}{Z'\bar N_{22}(\rht)}\; ,\\
\alpha &\equiv& \frac{3-\krho}{\krho-1}.
\end{eqnarray}
If the dark gas corresponds to a relatively small fraction of the total extinction
($x\ll 1$), then
\beq
\fdark(\krho)-\fdark(\krho=1)\simeq \frac{x^2}{2\alpha}.
\eeq
For $0\leq\krho\leq\frac 32$, where $|\alpha|\geq 3$, the 
magnitude of the difference between the
actual dark-gas fraction and that found in the text, where $\krho$ is taken to be unity,
is $\leq x^2/6$. This is typically small: For $\davdg <1$ and $Z'\bar N_{22}\simeq 1.5$,
as in the text, we have $x<0.5$ and $|\fdark(\krho)-\fdark(\krho=1)|<0.04$.
For the case $\bar N(\rht) \ll 4\dnsdg$ ($x\gg 1$) we see from equation
(\ref{eq:appenafdg}) that $f_{\rm DG} \sim 1$.

\section{Analytic Treatment of $A_V{\rm (H_2)}$}
\label{appen:h2form}

The analytic treatment of the H/${\rm H_2}$ transition assumes that 
${\rm H_2}$ is formed
on interstellar dust grains with an effective formation rate coefficient 
of ${\cal R}= 3\times 10^{-17}$ cm$^3$ s$^{-1}$,
and is destroyed by FUV photodissociation with an 
unshielded rate 
of $I_0=6\times 10^{-11}$ s$^{-1}$
in the \citet{habing1968}
field ($1.6 \times 10^{-3}$ erg cm$^{-2}$ s$^{-1}$)
or $I_0'=1.02\times 10^{-10}$~s\e\ in 
the \citet{draine1978} field, which is 1.7 times stronger.
In order to simulate an isotropic field 
incident from 2$\pi$
steradians on an optically thick slab, we assume a 1D FUV flux incident 
at an angle of 60 degrees
with the normal to the slab.  Consequently, to penetrate to a (normal) 
distance 
$x$ into the slab,
a photon needs to travel $2x$.     
In a steady state in which the photodissociation of molecular hydrogen is
balanced by the formation on dust grains, we have
\begin{equation}
G_0' I_0' f_s n_{{\rm H_2}} e^{-2b_{\rm H_2}A_V} = {\cal R}Z' n n_{\rm {HI}},
\label{eq:h2form}
\end{equation}
where $G_0'$ is the energy density of the dissociating radiation field in units of
the \citet{draine1978} field,
$f_s$ is the H$_2$ self-shielding factor (see below), $n_{{\rm H_2}}$ 
is the number density of molecular hydrogen,
$b_{\rm H_2}$ is the dust attenuation factor that accounts for 
differences in attenuation between the
visible and the FUV and also approximately includes the effects of 
FUV scattering,  $A_V$ is the
visual extinction into the slab normal to the surface
(see eq. \ref{eq:ns}), 
$n$ is the hydrogen 
nucleus number density,
and $n_{\rm {HI}}$ is the atomic hydrogen number density.   
In this Appendix and the following one, we are treating the gas in the clumps, where most
of the mass resides.  Therefore, there
is no need to distinguish the density $n$ from the local clump value $n_c$.
The factor of 2 
in the exponential follows from
the factor of 2 longer pathlength due to the oblique incidence of the 
FUV flux (see above).
We assume that
the surface area of grains increases proportionally to the increase 
in metallicity $Z'$, so that
both the H$_2$ formation rate coefficient and the FUV extinction 
increase linearly with $Z'$.

In order to obtain a relatively simple analytic approximation to the 
variation of $n_{{\rm H_2}}$ with depth 
into the slab, we require that the self-shielding factor be approximated 
by a power law with the form
\begin{equation}
f_s =  \left[ {N_1 \over 2N_{{\rm H_2}}}\right]^d\; ,
\label{eq:fsd}
\end{equation}
where $N_1$ is a constant that is determined by comparison with
detailed numerical calculations (see below), $N_{{\rm H_2}}$ is the H$_2$ column density
measured normal to the cloud surface, 
and the factor 2 is due to the longer pathlength associated with the 60 degree
incidence angle.

We wish to find an analytic expression for $A_V(\rht)$, 
where $A_V(\rht)$ is defined
as the characteristic value of the (normal) visual extinction to the point in the slab 
at which the hydrogen makes the transition from atomic to molecular form.
We define this to be at the point
where $n_{{\rm H_2}}= 0.25 n$ and 
$n_{{\rm HI}}= 0.5 n$.  In other words,
$A_V(\rht)$ is where the hydrogen is half 
atomic and half molecular.\footnote{Note that the simple
power law expression given in equation (\ref{eq:fsd}) 
need only apply for shielding
columns $N_{{\rm H_2}}\sim 10^{18-20}$ cm$^{-2}$ that correspond to typical
H$_2$ columns at positions in the slab with $A_V$ somewhat  less than or equal
to $A_V(\rht)$.}  
We denote the corresponding column densities by 
$N(\rht)$;
for example, $N_{{\rm H_2}}(\rht)$ is the column of H$_2$ from the surface of the cloud
at $R_{tot}$ to $R_{\rm H_2}$, or the column of H$_2$ to a depth given by $A_V(\rht)$.

Substituting $n_{{\rm H_2}}= 0.25 n$ and $n_{{\rm HI}}= 0.5 n$ into 
equation (\ref{eq:h2form}) we obtain for the
H$_2$ column $N_{{\rm H_2}}(\rht)$ corresponding to $A_V(\rht)$
\begin{equation}
N_{{\rm H_2}}(\rht)=0.5N_1 \left({{G_0' I_0'}\over {2{\cal R}Z'n}}\right)^{1/d} 
    e^{-{{2b_{\rm H_2}A_V(\rht)} / d}}\; .
\label{eq:n2p}
\end{equation}

We can also multiply equation (\ref{eq:h2form}) by $dx$ and 
integrate into the slab to obtain an
expression for $N_{{\rm H_2}}$ as a function of $N$ or $A_V$ 
into the slab.  We obtain the general solution up to the
point $N_{{\rm H_2}}(\rht)$ or $A_V(\rht)$  where the gas is half atomic and half molecular.  
In order to simplify this integration, we assume that 
$n_{{\rm HI}} = c_1n$ and is constant for $N < N_{{\rm H_2}}(\rht)$.  Note that $c_1$ varies 
from 1 at small column
to 0.5 as we approach $A_V(\rht)$,
and it can be adjusted
to best match the numerical code results.   
Since the solution heavily depends on what happens
near $A_V(\rht)$, we 
expect $c_1\simeq 0.5$; in addition, we expect the density to be $n\simeq n(R_{\rm H_2})$.  
For $d<1$  (which we show later 
is always the case) we obtain:
\begin{equation}
N_{{\rm H_2}}= 0.5\left[{{(1-d)c_1{\cal R}Z'nN_0}\over {b_{\rm H_2}G_0'I_0' N^{d}_1}}\right]^{1/(1-d)}
\left(e^{2b_{\rm H_2}A_V} -1\right) ^{1/(1-d)}.
\label{eq:n2}
\end{equation}
Substitution of $A_V=A_V(\rht)$ into this equation yields 
another expression for $N_{{\rm H_2}}(\rht)$
as a function of $A_V(\rht)$.

Equating the two expressions for $N_{{\rm H_2}}(\rht)$ (eqs.\,\ref{eq:n2p} and \ref{eq:n2}), 
we obtain 
a trancendental equation
whose solution gives $A_V(\rht)$:
\begin{equation}
\left[{{(1-d)c_1{\cal R}Z'nN_0}\over {b_{\rm H_2}G_0'I_0'  N^{d}_1}}\right]^{1/(1-d)}
\left(e^{2b_{\rm H_2}A_V(\rht)} -1\right) ^{1/(1-d)}=N_1 \left({{G_0' I_0'}
\over {2{\cal R}Z'n}}\right)^{1/d} e^{-{{2b_{\rm H_2}A_V(\rht)} / d}}\; .
\end{equation}
We can obtain analytic solutions in the two limits of small $A_V(\rht)$ and large $A_V(\rht)$.
In the limit $A_V(\rht) \ll [2b_{\rm H_2}]^{-1} \sim 0.25$ (see below), which corresponds to
no appreciable  dust extinction of FUV and entirely self-shielding by ${\rm H_2}$, we 
obtain
\begin{equation}
A_V(\rht) \simeq {1\over {(1-d)c_1 }}\left({{G_0' I_0'} \over {2{\cal R}Z'n}}\right)^{1/d}
\left({N_1 \over N_0}\right) \; .
\end{equation}
In the limit $A_V(\rht) \gg [2b_{\rm H_2}]^{-1} \sim 0.25$, which corresponds to significant dust
extinction as well as self-shielding, we obtain
\begin{equation}
A_V(\rht) \simeq {d \over {2b_{\rm H_2}}}\ln \left[{{2b_{\rm H_2} N_1} \over{(1-d) c_1  N_0}}
    \left({{G_0' I_0'}\over{2{\cal R}Z'n}}\right)^{1/ d}\right]\; .
\end{equation}
These limiting solutions correspond to 
$G_0'/n \ll 0.019$ cm$^3$ and $G_0'/n \gg 0.019$ cm$^3$, respectively.
The two limiting solutions can be combined in a simple expression that preserves
the limiting solutions, and smoothly transitions from one to the other when
$A_V(\rht)  \sim [2b_{\rm H_2}]^{-1} \sim 0.25$:
\begin{equation}
A_V(\rht) \simeq {d \over {2b_{\rm H_2}}}\ln \left[ 1 + {{2b_{\rm H_2} N_1} 
     \over{d(1-d)c_1 N_0}}    \left({{G_0' I_0'}
\over{2{\cal R}Z'n}}\right)^{1/d}\right]
\end{equation}

Our numerical code uses the Meudon code to solve for the dust extinction 
and ${\rm H_2}$ self-shielding
that leads to the H/${\rm H_2}$ transition in a slab.   From the code results 
for $Z'=1$, we find that for our range of 
$G_0'/n$ (which extends from $4.2 \times 10^{-3}$ cm$^3$ to 0.3 cm$^3$),
$A_V(\rht) \sim 0.05 - 1$ and $N_{{\rm H_2}}(\rht) \sim 10^{19} - 10^{20}$ cm$^{-2}$.   We fit the
self shielding for the critical range $10^{18}$ cm$^{-2} < N_{{\rm H_2}} 
< 10^{20}$ cm$^{-2}$ with
 $N_1= 3.6 \times 10^{12}$ cm$^{-2}$ and $d = 0.57$. We fit 
the dust shielding of the FUV with $b_{\rm H_2} \simeq 2$.  In the limited allowed
range $0.5 < c_1 < 1$, we find that $c_1= 0.5$ best
fits the numerical results, as expected from the above discussion. Therefore, we obtain the analytic 
expression
\begin{equation}
A_V(\rht) \simeq 0.142 \ln \left[ 1 + 5.2 \times 10^3 Z' 
   \left({G_0' \over {Z' n}}\right)^{1.75}\right],
\label{eq:avh2}
\end{equation}
where $n$ is in cm$^{-3}$ and $G_0'$ references the Draine field.  
Over the relevant range
$4.2 \times 10^{-3} < G_0'/n < 0.3$
(which corresponds 
to $ 0.05 < A_V(\rht)< 1.1$), and for $Z' = 0.5$, 1, and 1.9, 
we find that this 
analytic solution matches the numerical code result to better than 5\%.

\section{Analytic Treatment of $A_V{\rm (CO)} $}   
\label{appen:coform}

In our standard runs with relatively low cosmic-ray ionization rates (primary rates 
$\sim 2 \times 10^{-17}$ s$^{-1}$),
the chemistry of CO from the surface of the cloud to $A_V{\rm (CO)}$ 
is dominated by the formation of OH on
grain surfaces followed by a chemical chain that leads to CO.    
The OH abundance is determined by
equating the formation of OH on grain surfaces 
(O\ + \ gr\   $\rightarrow$ \ OH)
to the FUV photodissociation of OH
(OH\ + \  $h\nu$\  $\rightarrow$ \  O \ + \ H):
\begin{equation}
\gamma _1  \xo n^2 = I_{{\rm OH}}' G_0' e^{-2b_{{\rm OH} }A_V} 
     \xoh n,
\label{eq:OHform}
\end{equation}
where $n$ is the hydrogen nucleus density, 
$\gamma _1 \simeq 5 \times 10^{-17}Z'$ cm$^3$ s$^{-1}$
\citep{hollenbach2009}
is the effective rate coefficient for O atoms colliding and sticking 
to dust grains (the dust cross section
per H nucleus is contained in $\gamma _1$), $\xo n = n({\rm O})$ is the density 
of O atoms in the gas, $I_{{\rm OH}}' = 3.5 \times
10^{-10}$ s$^{-1}$ 
\citep{roberge1991,woodall2007}
is the (unshielded) photodissociation rate of OH in 
the local (Draine) 
interstellar field, $G_0'$ is the ratio
of the incident FUV field to the Draine field, 
$b_{{\rm OH}}= 1.7$ 
is the factor that, multiplied by $A_V$, gives
the effective grain optical depth in the FUV to photons that 
dissociate OH, $\xoh n = n({\rm OH})$ is the OH number density
in the gas, and the factor of 2 in the exponential reflects the 
assumption that a diffuse field incident
on a cloud can be approximated by a 1D flux incident at an angle of 
60 degrees to the normal.  Note that
we assume that every atomic O that strikes a grain sticks to the grain, 
reacts with an H atom, and comes
off the grain as OH.  Equivalently, it might react with another H atom, form 
${\rm H_2O}$ on the grain surface
and photodesorb as either OH or as ${\rm H_2O}$, which 
immediately photodissociates to OH.
We derive from eq. (\ref{eq:OHform}):
\begin{equation}
\xoh = {{\gamma _1 \xo n}
            \over {I_{{\rm OH}}'G_0' e^{-2b_{{\rm OH}}A_V}}}
\label{eq:OHden}
\end{equation}

Once OH is formed the chemistry proceeds as follows:

OH\ \ + \ \ C$^+$\ \  $\rightarrow$ \ \ CO$^+$ \ \ + \ \ H.

CO$^+$\ \ + \ \ H$_2$\ \  $\rightarrow$ \ \ HCO$^+$ \ \ + \ \ H.

HCO$^+$\ \ + \ \ e\ \  $\rightarrow$ \ \ CO \ \ + \ \ H.

CO\ \ + \ \ $h\nu$\ \  $\rightarrow$ \ \ C \ \ + \ \ O.

\noindent The first reaction dominates the formation of CO$^+$; the 
second reaction
dominates the destruction of CO$^+$ and the formation of HCO$^+$; the 
third reaction
dominates the destruction of HCO$^+$ and the formation of CO; the 
last reaction
dominates the destruction of CO.   As a result, every CO$^+$ that is 
formed by the first reaction
results in a formation of a CO by the third reaction.   We can then 
equate the formation rate
of CO (the first reaction) with the destruction rate of CO 
(the last reaction):
\begin{equation}
\gamma_2 \xoh\xcp n^2 = I_{{\rm CO}}'
                G_0' f_{{\rm CO}} e^{-2b_{{\rm CO}}A_V} \xco n,
\label{eq:COform}
\end{equation}
where $\gamma _2 = 2.9 \times 10^{-9}(T/300\ {\rm K})^{-0.33}$ cm$^3$ s$^{-1}$ 
(Dubernet et al.\ 1992; E. Herbst 2006 private communication\footnote{see
also 
{http://www.physics.ohio-state.edu/$\sim$eric/research.html}})
is the
rate coefficient for OH reacting with C$^+$ (first reaction), 
$\xcp n$ is 
the C$^+$ gas phase
number density, 
$I_{{\rm CO}}'= 2.6 \times 10^{-10}$ s$^{-1}$ 
\citep{visser2009}
is the 
unshielded photodissociation rate
of CO in the Draine field, 
$f_{{\rm CO}}$ is the shielding factor caused by 
CO (self-shielding) and
H$_2$, the exponential factor is shielding of CO by dust, $b_{\rm CO}= 3.2$, 
and $\xco n$ is the
number density of CO.

Near $A_V(\rco)$, the CO photosphere, we find by 
fitting to \cite{visser2009} that
\begin{equation}
f_{\rm CO} = c_{\rm CO} 
\left[{{2N_\co} \over {10^{16}\ {\rm cm}^{-2}}}\right]^{-0.60},
\label{eq:COshield}
\end{equation}
where $c_{\rm CO} = 4.4 \times 10^{-2}$ and where the factor of 
2 applies because
$N_\co$ is the column of CO along a line of sight perpendicular 
to the surface, whereas
our isotropic field is roughly equivalent to a 1D flux incident at an angle of 60
degrees to the surface.  The flux then traverses a column 2$N_\co$ to 
the point in
question (see the discussion of  $A_V(\rht)$ in Appendix \ref{appen:h2form}).

We substitute equation (\ref{eq:COshield}) and equation (\ref{eq:OHden}) 
into equation (\ref{eq:COform}) and integrate both sides of the equation
over $dz$, the perpendicular depth into the cloud, until we reach 
the CO photosphere,
$N_\co= 2\times 10^{16}$ cm$^{-2}$ at $A_V(\rco)$.  
Here we use $ndz = N_0 dA_V$ with
$N_0 = 1.9 \times 10^{21}/Z'$ cm$^{-2}$ and $\xco ndz= dN_\co$:
\begin{eqnarray}
\int _0^{A_V(\rco)} 
\left[ {{\gamma _1 \gamma _2 \xo\xcp n^2}\over 
         {I_{{\rm OH}}' G_0'}}\right] e^{2(b_{{\rm OH}}+b_{{\co}})A_V} dA_V=~~~~~~~~~~~~~~~~~~~~~~~~~~~~~~~~~~~~\\
~~~~~~~~~~\int _0^{2\times 10^{16}\ {\rm cm}^{-2}} I_{\rm CO}' G_0' 
       \left[ {{2N_\co}\over {10^{16}\ {\rm cm}^{-2}}}\right]
       ^{-0.60} dN_\co.
\end{eqnarray}
Assuming that  
the density is constant from the surface to $A_V(\rco)$ and that the carbon is entirely
C$^+$ ($\xcp=\xc$),
we solve for
 $A_{V}({\rm CO})$:
\begin{equation}
A_V(\rco) = {1\over {2(b_{\rm OH} + b_{\rm CO})}} \ln
\left[ {{4.35 \times 10^{16} 
c_{\rm CO} (b_{\rm OH} +b_{\rm CO})I_{\rm OH}' I_{\rm CO}' G_0'^2}\over {
   N_0 \gamma _1 \gamma _2 x_{\rm O} x_{\rm C} n^2}} + 1 \right],
\label{eq:avco}
\end{equation}
where $x_{\rm O} \simeq 3.2 \times 10^{-4}Z'$ is the gas phase abundance of 
atomic O, $x_{\rm C} \simeq
(1.6 \times 10^{-4}) Z'$ is the gas phase abundance of atomic carbon, 
 and $Z'$ is the metallicity relative to solar. We find this analytic solution
is good to within 15\% except for the lowest $Z'$ and $G_0'$ model 
($Z' = 0.5$, $G_0' = 3$), where 
the fit is good to within 25\%.

\newpage
\centerline{Figures}
\plotone{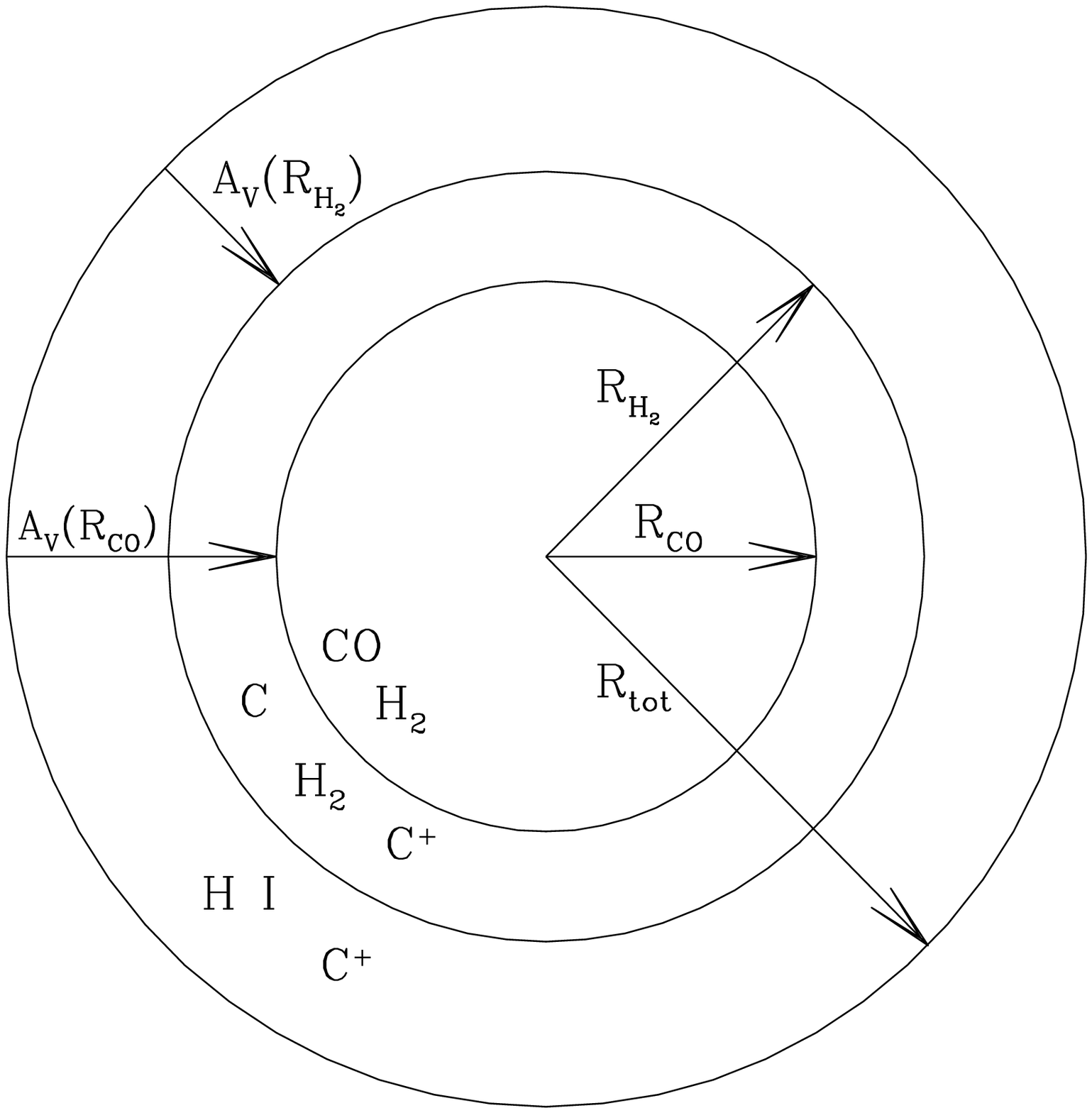}
\figcaption[avdiag.eps]{Illustration of a model cloud showing
the radius $\rco$  of the CO core, the radius $\rht$
where $2n_{{\rm H_2}} = n_{{\rm HI}}$ (equal mass density
in H atoms and H$_2$ molecules), and $R_{\rm tot}$ the total cloud radius.
Within $R < \rco$, gas is mainly CO and ${\rm H_2}$. Within
the range $\rco < R < \rht$, gas is mainly ${\rm H_2}$ whereas the
gas phase carbon is mainly C and C$^+$.
Within the range $\rht < R < R_{\rm tot}$ gas is mainly 
\ion{H}{1} whereas the gas phase carbon is mainly C$^+$. 
$A_V(\rht)$ is the optical depth measured from the
outer radius to $\rht$ and $A_V(\rco)$ is the optical depth
measured from the outer radius to $\rco$.
\label{fig:avdiag}}

\plotone{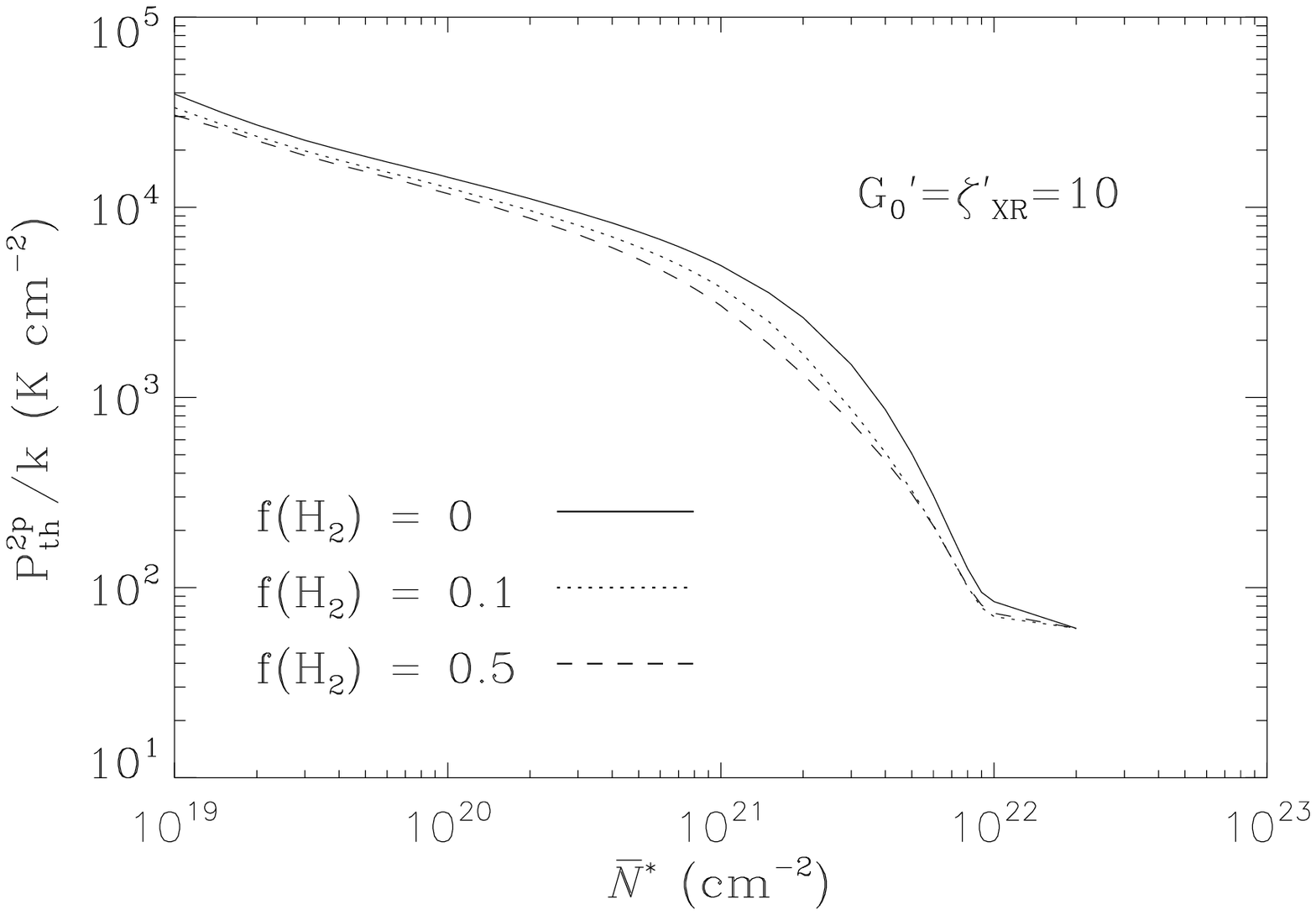}
\figcaption[paveplt.eps]{Average two-phase thermal pressure 
$P_{\rm th}^{\rm 2p} =  (P_{\rm min}P_{\rm max})^{1/2}$ as a function
of total column density from the cloud surface 
$\bar{N}$ and molecular
fraction $f({\rm H_2}) = 2{N}_{\rm H_2}/{N}$. 
The FUV radiation field and
soft X-ray/EUV ionization rates are a factor of 10 times higher than 
local interstellar medium values. Curves are shown for 
$f({\rm H_2}) = 0$ ({\em solid curve}),
$f({\rm H_2}) = 0.1$ ({\em dotted curve}), and
$f({\rm H_2}) = 0.5$ ({\em dashed curve})
\label{fig:paveplt}}

\plotone{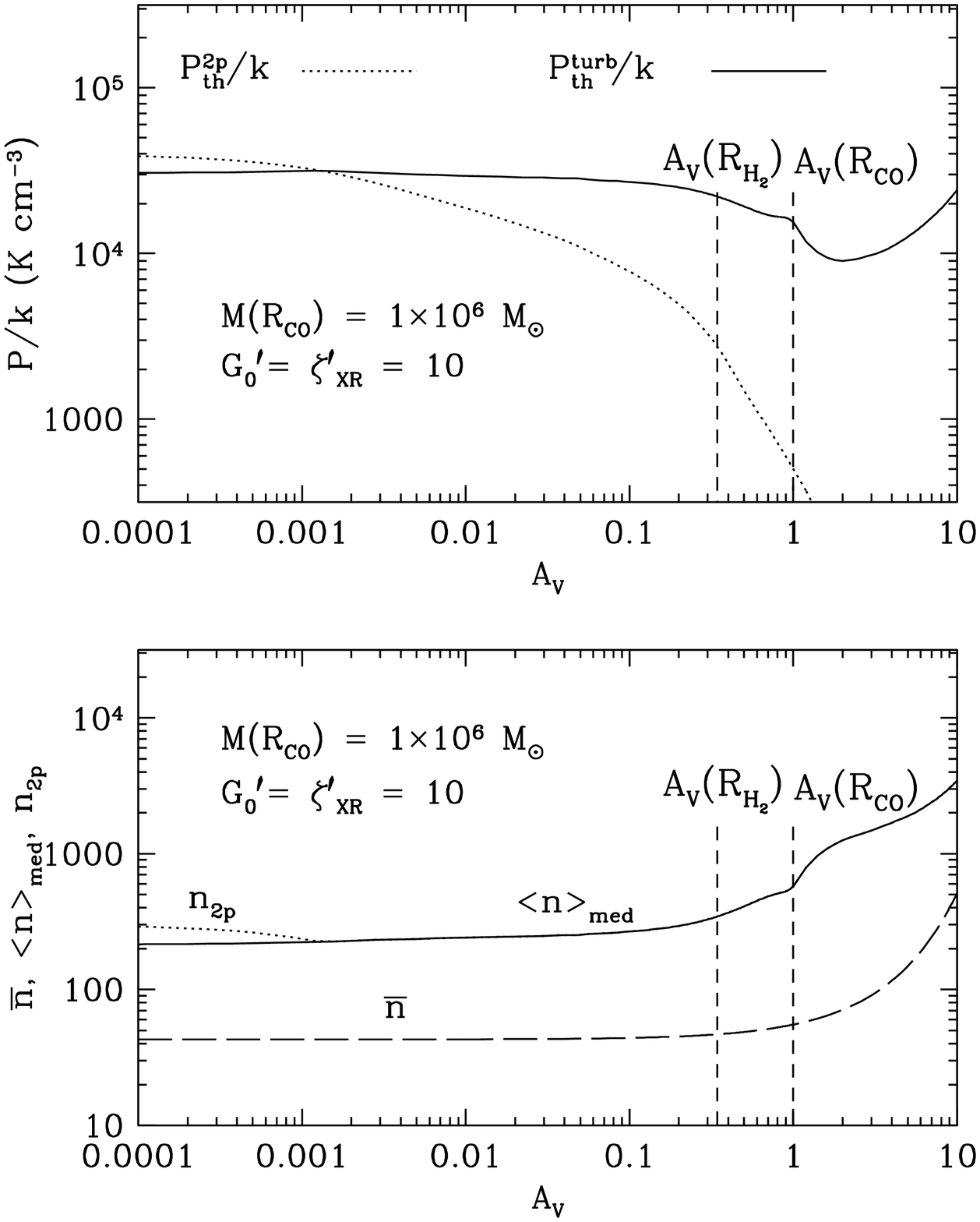}
\figcaption[Preslg.eps]{{\em Top panel:} Thermal pressure as a
function of optical depth into the cloud for 
$M(\rco) = 1\times 10^{6}$ $M_{\odot}$, $\bar N_{22} = 1.5$, 
$Z'=1$, 
and incident radiation
field $G_0' = \zeta_{\rm XR}' =  10$. Curves are shown for two-phase thermal
pressure 
$P_{\rm th} ^{\rm 2p}/k = x_t n_{\rm 2p} T/k$ ({\em dotted curve})
and for turbulent thermal pressure
$P^{\rm th}_{\rm turb}=x_t \langle n 
\rangle_{\rm med}T/k$ ({\em solid curve}).
Optical depths at $A_V(\rht)$ [$n_{{\rm H_2}}/n=0.25$] and
$A_V(\rco)$ ($\tau_{\rm CO} =1$) are indicated by vertical
dashed lines. Thermal pressure in the turbulent medium 
dominates at $A_V \ga 0.001$.
{\em Bottom panel:} Density as a function of optical depth into 
the cloud for 
$M(\rco) = 1\times 10^{6}$ $M_{\odot}$, $\bar N_{22}=1.5$,
$Z'=1$, and incident radiation
field $G_0' = \zeta_{\rm XR}' = 10$. Curves are shown for two-phase density 
$n_{\rm 2p}$, mass-weighted median density $\langle n \rangle_{\rm med}$
in a turbulent density distribution, and volume-averaged density
$\bar{n}$. The two-phase density distribution is shown only where
two-phase pressure dominates ($A_V \la 0.001$). The local 
(model) density 
$n_c=n_{\rm 2p}$ where two-phase pressure dominates
and $n_c = \langle n \rangle_{\rm med}$ where turbulent pressure dominates.
Optical depths at $A_V(\rht)$ [$n_{{\rm H_2}}/n=0.25$] and
$A_V(\rco)$ ($\tau_{\rm CO} =1$) are indicated by vertical
dashed lines. 
\label{fig:Preslg}}

\plotone{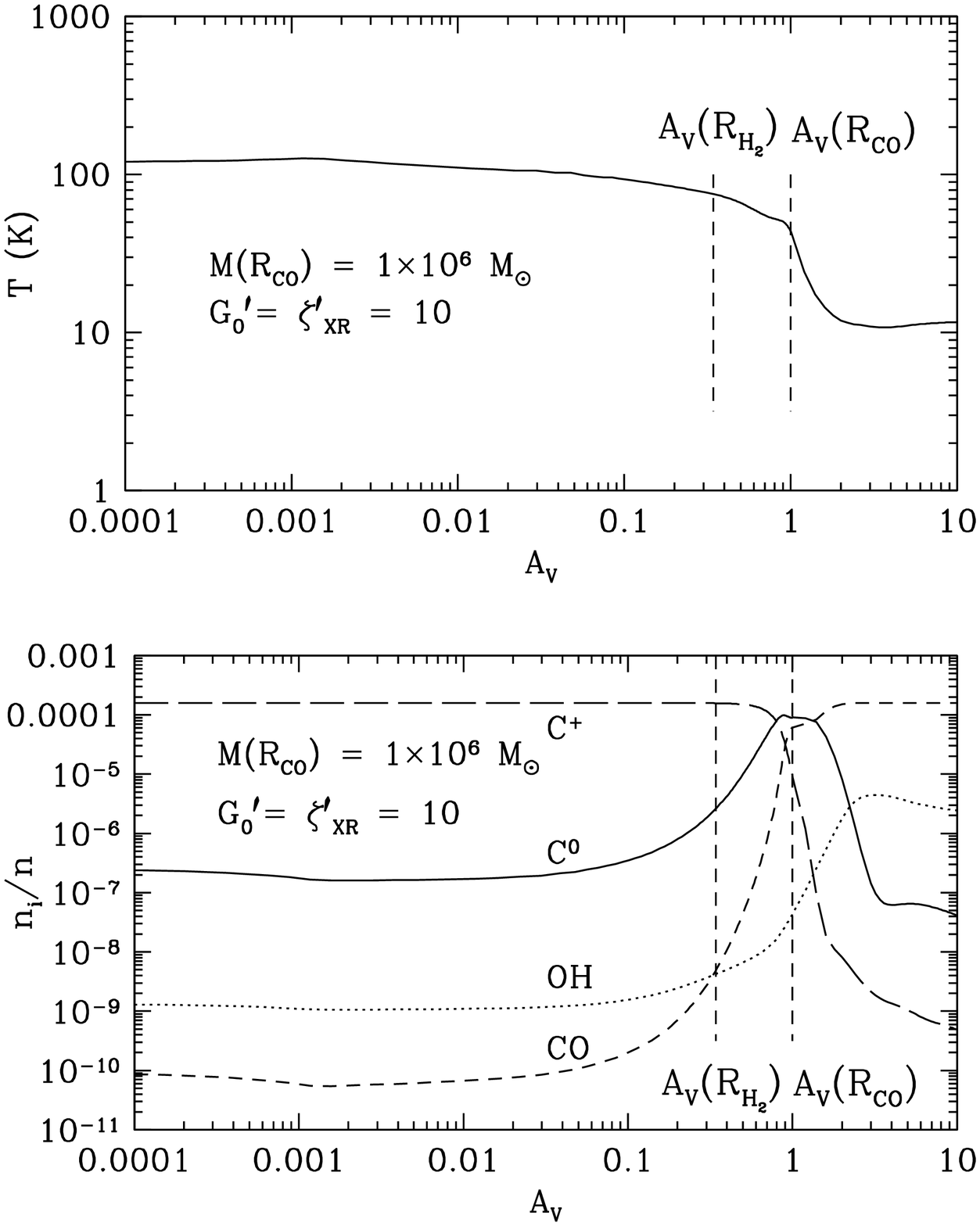}
\figcaption[templg.eps]{{\em Top panel:} Temperature 
as a function of optical depth into
the cloud for
$M(\rco) = 1\times 10^{6}$ $M_{\odot}$, $\bar N_{22}=1.5$,
$Z'=1$,  and incident radiation
field $G_0' = \zeta_{\rm XR}' = 10$.
{\em Bottom panel:} Abundances of
${\rm C^+}$ ({\em long-dash curve}), ${\rm C^0}$ ({\em solid curve}),
${\rm OH}$ ({\em dotted curve}), and ${\rm CO}$ ({\em short-dash curve})
as functions of optical depth into
the cloud for
$M(\rco) = 1\times 10^{6}$ $M_{\odot}$, $\bar N_{22}=1.5$,
$Z'=1'$, and incident radiation
field $G_0' = \zeta_{\rm XR}' = 10$.
Note that we do not include freeze out
of ${\rm H_2O}$ on grain surfaces, which would affect OH
abundances at 
$A_V > 3$.
\label{fig:templg}}

\plotone{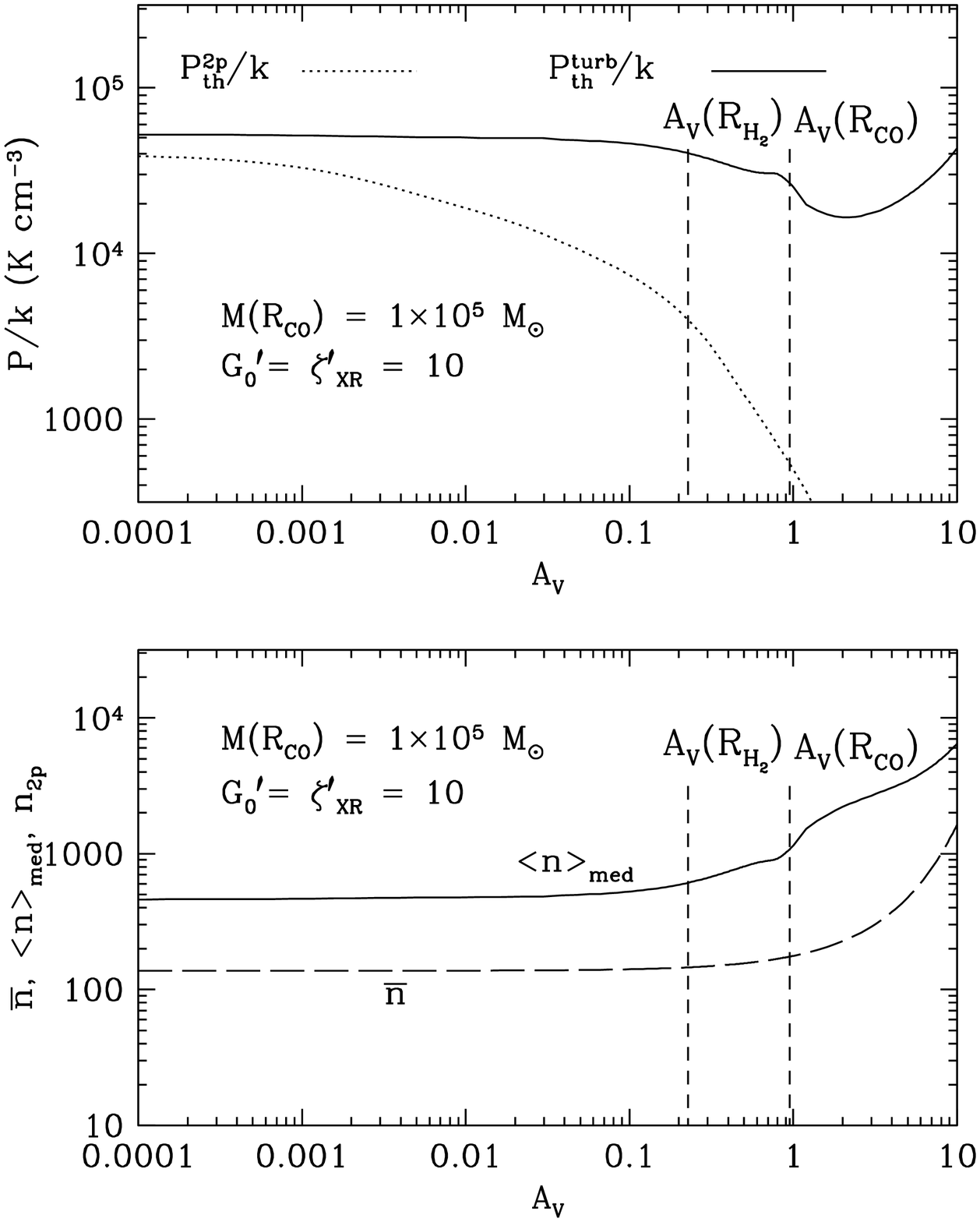}
\figcaption[Preslglm.eps]{{\em Top panel:} Thermal pressure as a
function of optical depth into the cloud for 
$M(\rco) = 1\times 10^{5}$ $M_{\odot}$, $\bar N_{22}=1.5$,
$Z'=1$,  and incident radiation
field $G_0' = \zeta_{\rm XR}' =  10$. Curves are shown for two-phase thermal
pressure 
$P_{\rm th} ^{\rm 2p}/k = x_t n_{\rm 2p} T/k$ ({\em dotted curve})
and for turbulent thermal pressure
$P^{\rm th}_{\rm turb}=x_t \langle n 
\rangle_{\rm med}T/k$ ({\em solid curve}).
Optical depths at $A_V(\rht)$ [$n_{{\rm H_2}}/n=0.25$] and
$A_V(\rco)$ ($\tau_{\rm CO} =1$) are indicated by vertical
dashed lines. Thermal pressure in the turbulent medium 
dominates at all $A_V$.
{\em Bottom panel:} Density as a function of optical depth into 
the cloud for 
$M(\rco) = 1\times 10^{5}$ $M_{\odot}$, $\bar N_{22}=1.5$,
$Z'=1$,  and incident radiation
field $G_0' = \zeta_{\rm XR}' = 10$. 
Curves are shown for 
mass-weighted median density $\langle n \rangle_{\rm med}$
in a turbulent density distribution, and volume-averaged density
$\bar{n}$. Turbulent pressure dominates at all $A_V$ and
thus no two-phase density is shown.
The local (model) density 
$n_c = \langle n \rangle_{\rm med}$.
Optical depths at $A_V(\rht)$ [$n_{{\rm H_2}}/n=0.25$] and
$A_V(\rco)$ ($\tau_{\rm CO} =1$) are indicated by vertical
dashed lines. 
\label{fig:Preslglm}}

\plotone{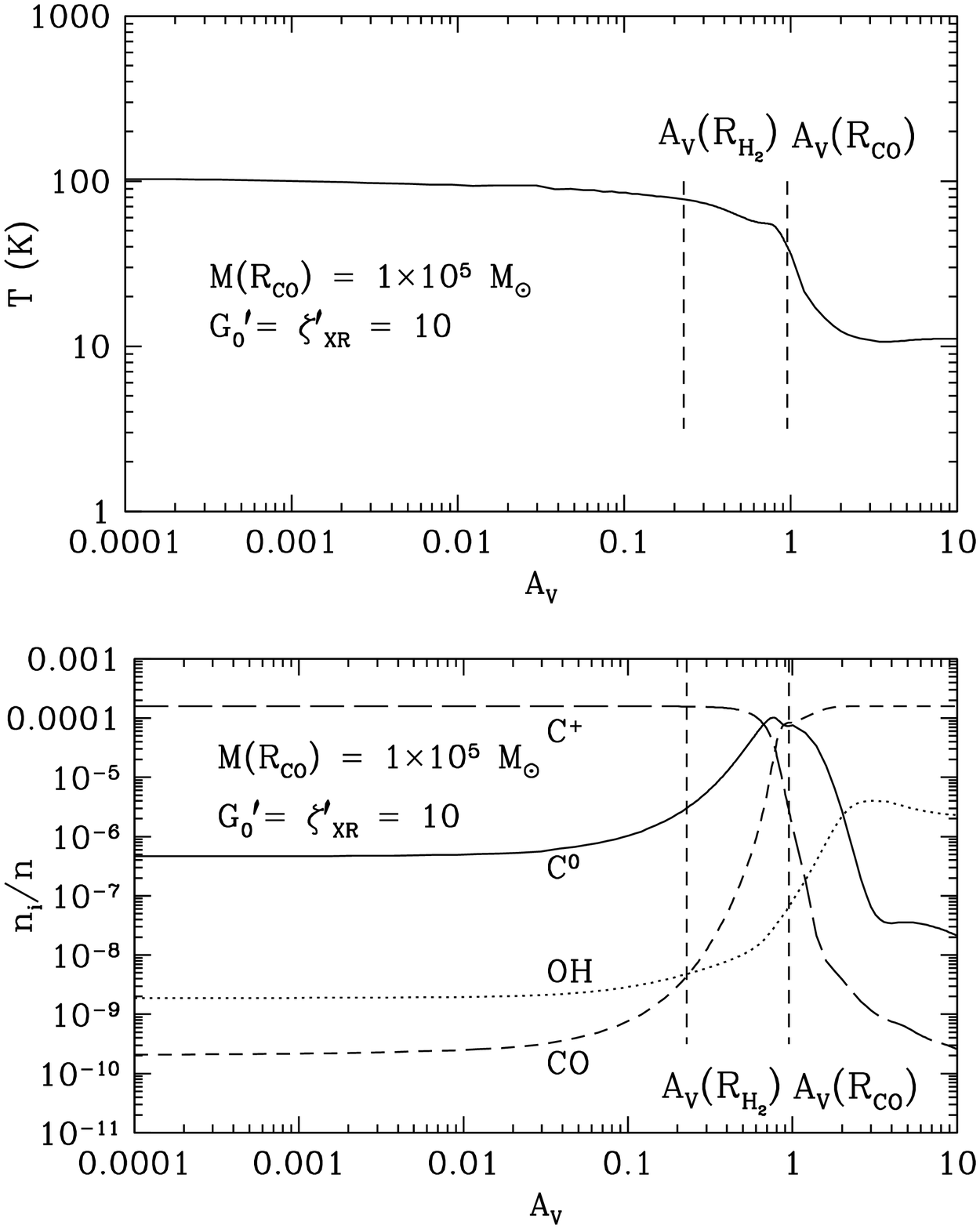}
\figcaption[templglm.eps]{{\em Top panel:} Temperature 
as a function of optical depth into
the cloud for
$M(\rco) = 1\times 10^{5}$ $M_{\odot}$, $\bar N_{22}=1.5$, 
$Z'=1$, and incident radiation
field $G_0' = \zeta_{\rm XR}' = 10$.
{\em Bottom panel:} Abundances of
${\rm C^+}$ ({\em long-dash curve}), ${\rm C^0}$ ({\em solid curve}),
${\rm OH}$ ({\em dotted curve}), and ${\rm CO}$ ({\em short-dash curve})
as functions of optical depth into
the cloud for
$M(\rco) = 1\times 10^{5}$ $M_{\odot}$, $\bar N_{22}=1.5$,
 and incident radiation
field $G_0' = \zeta_{\rm XR}' = 10$.
Note that we do not include freeze out
of ${\rm H_2O}$ on grain surfaces, which would affect OH
abundances at 
$A_V > 3$.
\label{fig:templglm}}

\plotone{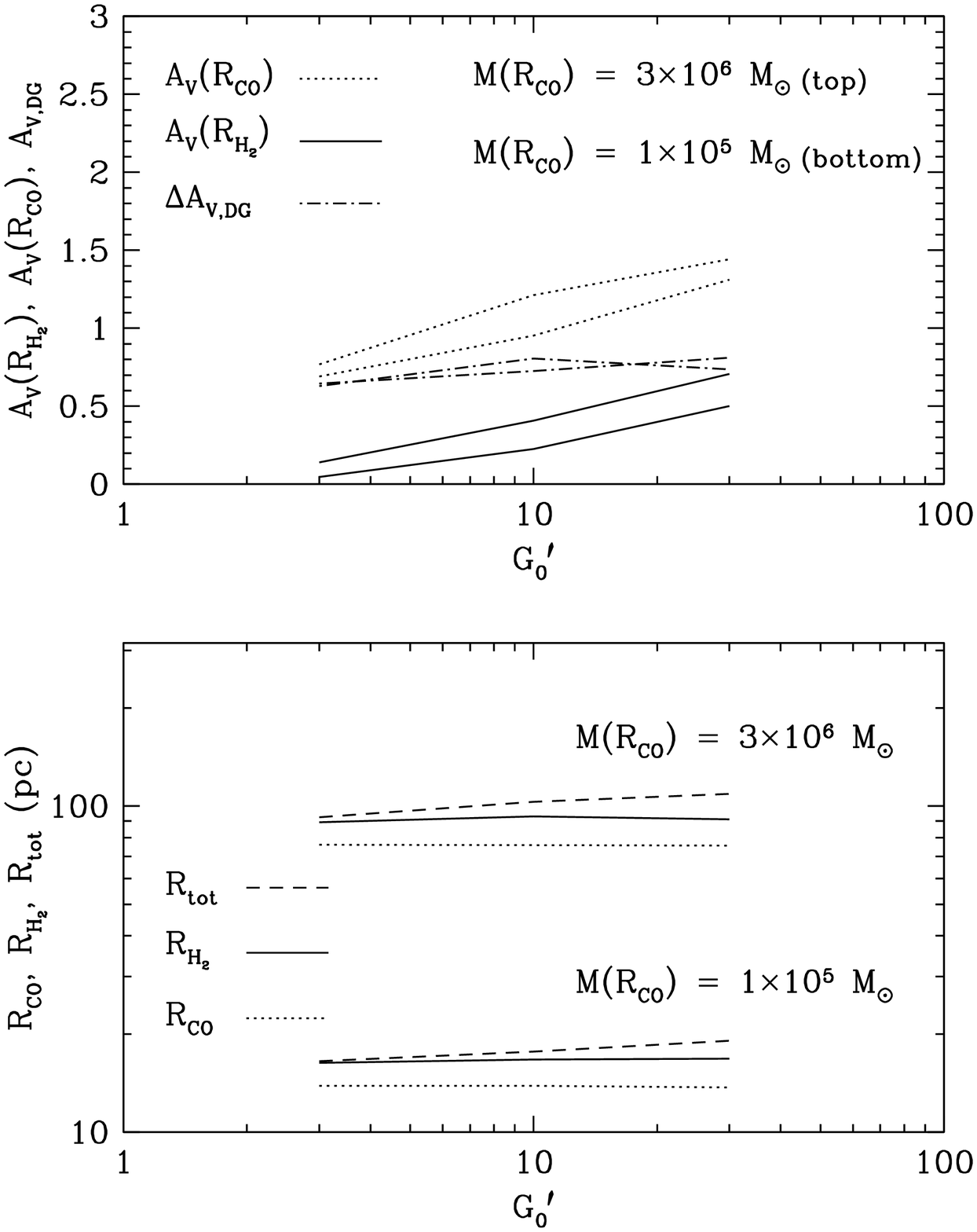}
\figcaption[Rout.eps]{{\em Top panel:} 
Optical depth as a function of cloud mass, $M(\rco)$, and incident
radiation field, $G_0'=G_0/1.7$. 
Curves are shown for $\bar N_{22}=1.5$, $Z'=1$, and two cloud masses,
$M(\rco) = 3\times 10^6$ $M_\odot$ and 
$M(\rco) = 1\times 10^5$ $M_\odot$.
The optical depth from the cloud surface 
$R_{\rm tot}$ to $\rco$ is $A_V(\rco)$ ({\em dotted curve})  and the 
optical depth
from the cloud surface 
$R_{\rm tot}$ to $\rht$ is $A_V(\rht)$ ({\em solid curve}). 
Also shown is $\davdg=A_V(\rco)-A_V(\rht)$ ({\em dash-dot curve})  
{\em Bottom panel:} Cloud 
radii $R_{\rm tot}$ ({\em dashed curve}), $\rht$ ({\em solid curve}), and
$\rco$ ({\em dotted curve}) are shown as functions of cloud mass $M(\rco)$ 
and incident radiation field 
normalized to the local interstellar field, 
$G_0'=G_0/1.7$.  
Curves are shown for $\bar N_{22}=1.5$, $Z'=1$,  and two cloud masses
$M(\rco) = 3\times 10^6$ $M_\odot$ and 
$M(\rco) = 1\times 10^5$ $M_\odot$.
The radius $\rht$ is where
half the nuclei are in ${\rm H_2}$ 
($n_{{\rm H_2}}/n_c = 0.25$), and
the radius $\rco$ is where $\tau_{\rm CO} =1$. 
\label{fig:Rout}}

\plotone{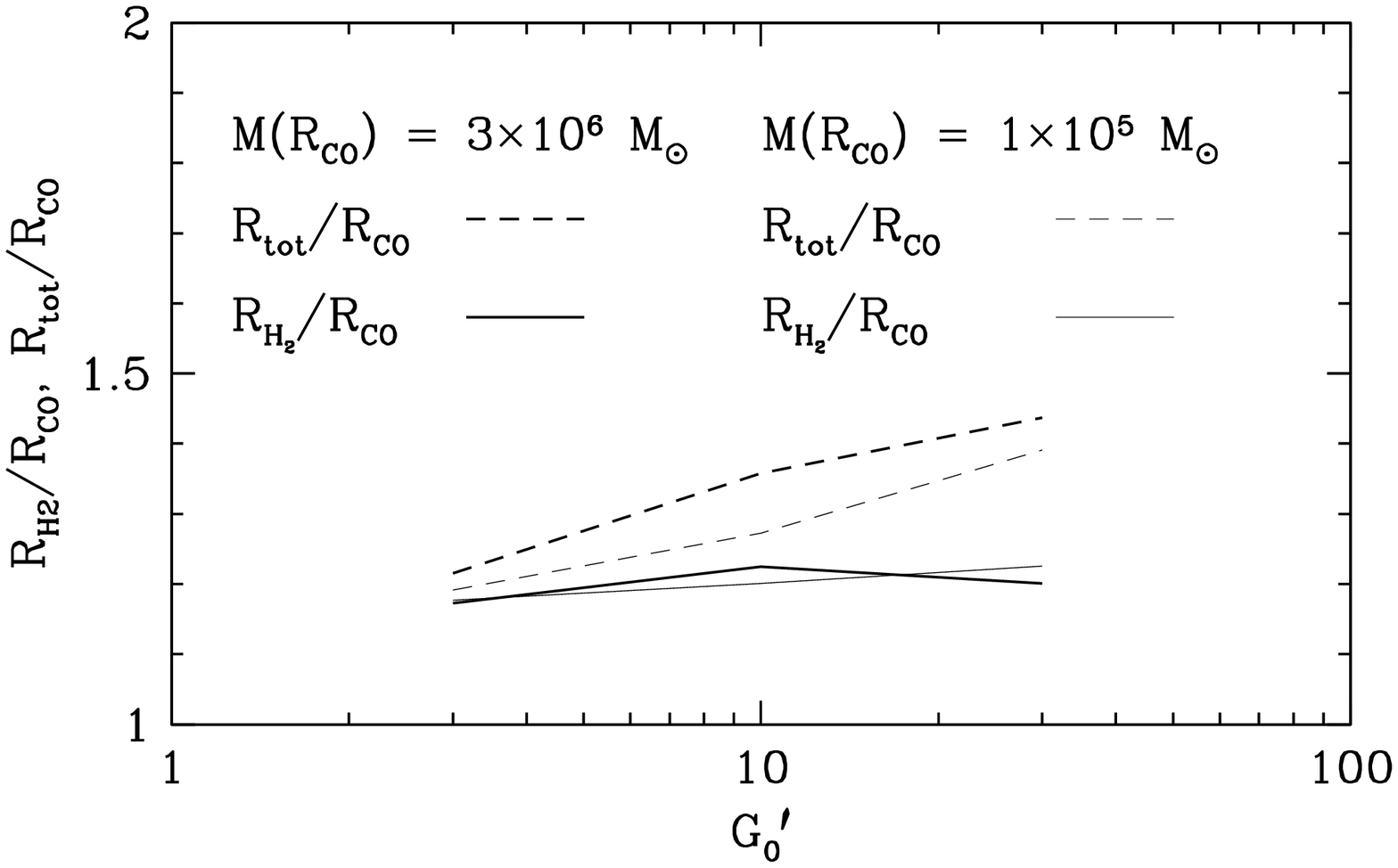}
\figcaption[Routrat.eps]{
Ratio of cloud radii $R_{\rm tot}/\rco$ ({\em dashed curve}), and
$\rht/\rco$ ({\em solid curve}) are shown as functions of cloud mass $M(\rco)$ 
and incident radiation field 
normalized to the local interstellar field, 
$G_0'=G_0/1.7$.  
Curves are shown for $\bar N_{22} = 1.5$, $Z'=1$,  and two cloud masses
$M(\rco) = 3\times 10^6$ $M_\odot$ ({\em thick curve})
and
$M(\rco) = 1\times 10^5$ $M_\odot$ ({\em thin curve}).
The radius $\rht$ is where
half the nuclei are in ${\rm H_2}$ 
($n_{{\rm H_2}}/n_c = 0.25$), and
the radius $\rco$ is where $\tau_{\rm CO} =1$. 
For $M(\rco) = 3\times 10^6$ $M_\odot$, $\rco = 75.4$ pc
and for $M(\rco) = 1\times 10^5$ $M_\odot$, 
$\rco = 13.8$ pc.
\label{fig:Routrat}}

\plotone{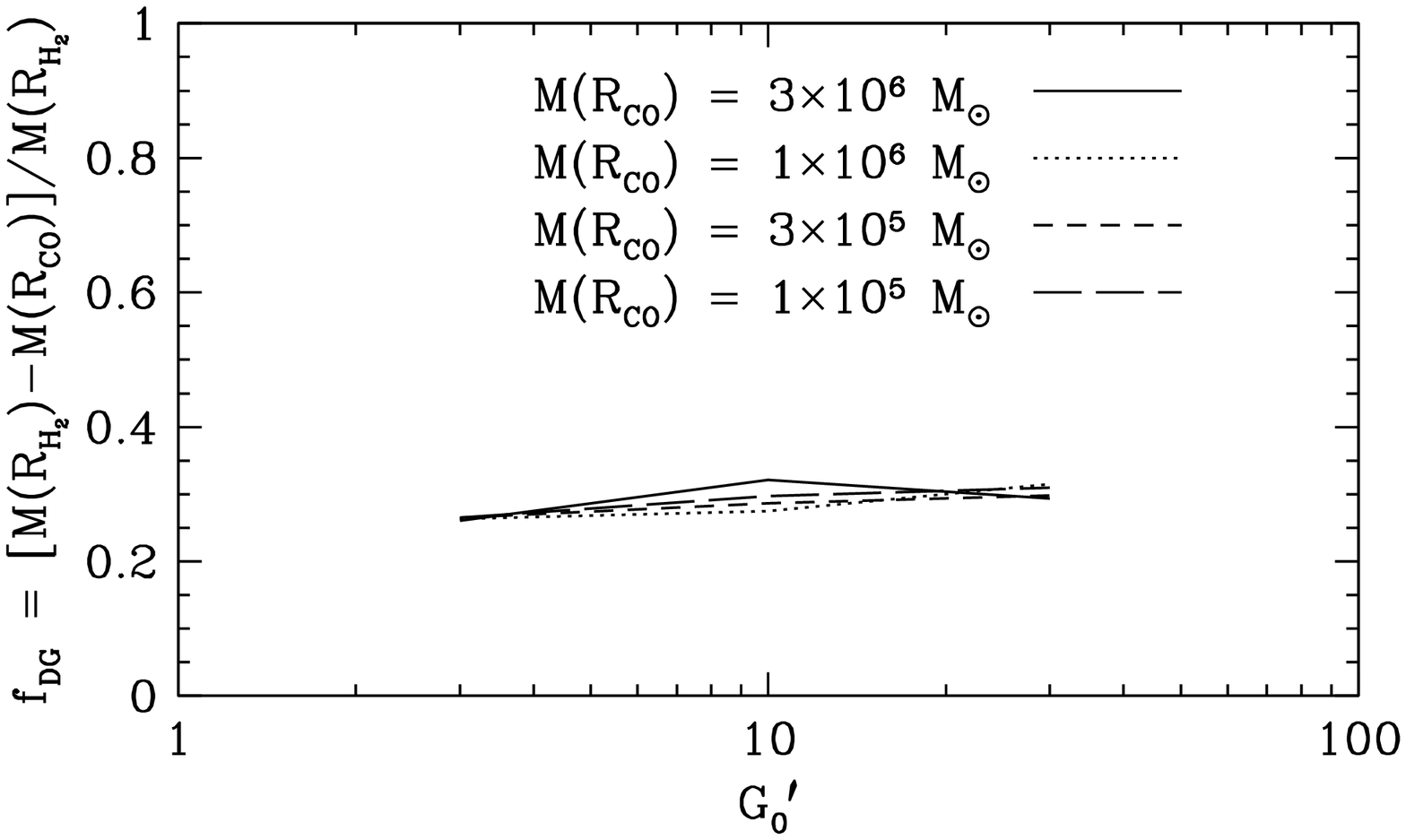}
\figcaption[ratio.eps]{
Dark gas fraction $f_{\rm DG} = [M(\rht)-M(\rco)]/M(\rht)$
versus incident radiation field
normalized to the local interstellar field, 
$G_0'=G_0/1.7$.  Curves are shown
for $\bar N_{22} =1.5$, $Z'=1$, and cloud masses 
$M(\rco)=3\times 10^{6}$ $M_{\odot}$ ({\em solid curve}),
$1\times 10^{6}$ $M_{\odot}$ ({\em dotted curve}),
$3\times 10^{5}$ $M_{\odot}$ ({\em short-dash curve}), and
$1\times 10^{5}$ $M_{\odot}$ ({\em long-dash curve}).
\label{fig:ratio}}

\plotone{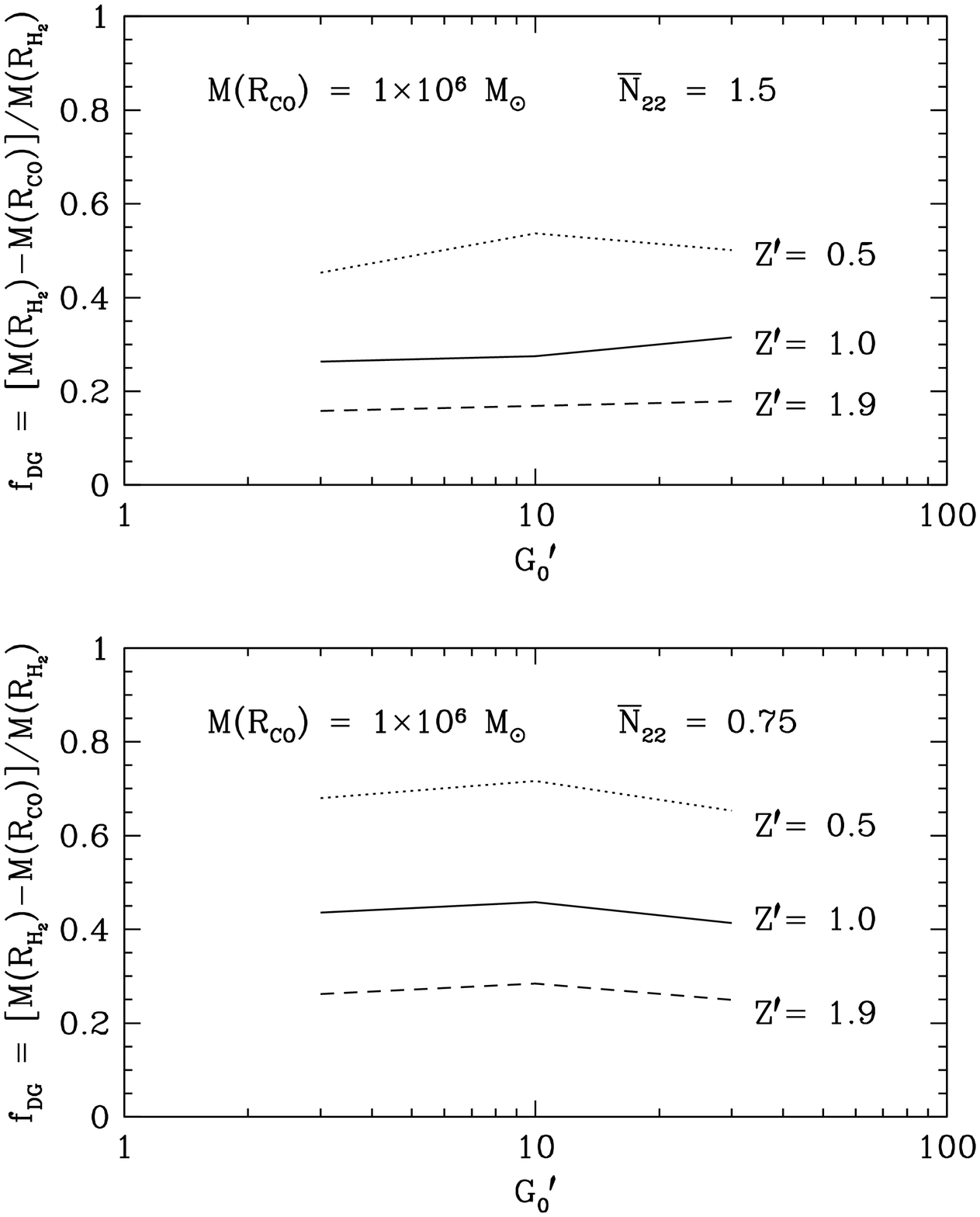}
\figcaption[fdglz2.eps]{{\em Top panel:} 
Dark gas fraction $f_{\rm DG} = [M(\rht)-M(\rco)]/M(\rht)$
versus incident radiation field normalized to 
the local interstellar field, 
$G_0'=G_0/1.7$.  Curves are shown
for constant mean column density $\bar{N}_{22} = 1.5$ and
metallicities $Z' = 1.9$ ({\em dashed curve}), $Z' = 1$ ({\em solid curve}),
$Z' = 0.5$ ({\em dotted curve}). {\em Bottom panel:} Same as top
panel for $\bar{N}_{22}=0.75$.
\label{fig:fdglz2}}

\plotone{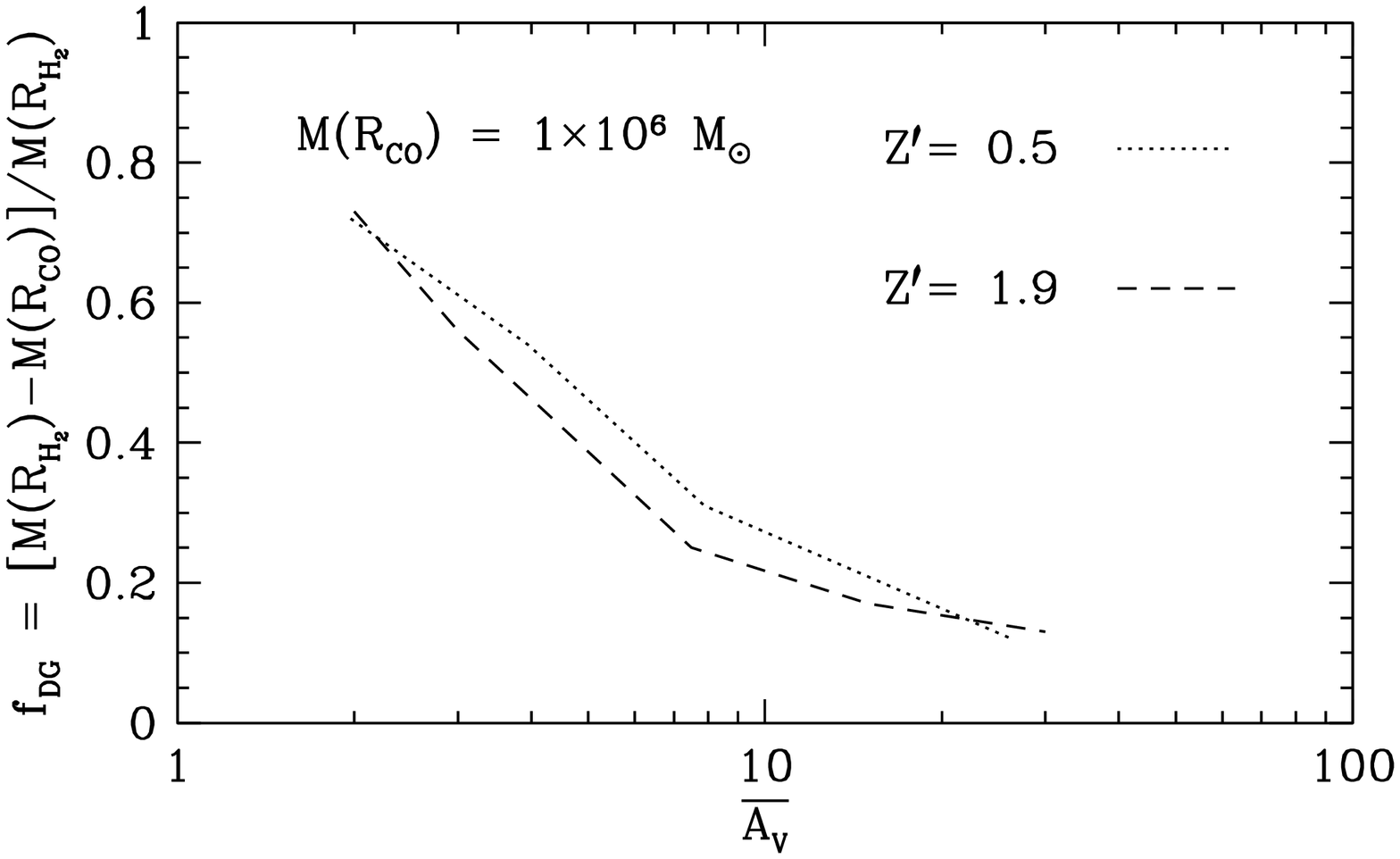}
\figcaption[avmean.eps]{{\em Top panel:} 
Dark gas fraction $f_{\rm DG} = [M(\rht)-M(\rco)]/M(\rht)$
as a function of the mean visual extinction through the cloud
 $\bar{A}_V=5.26 Z'\bar{N}_{22}$.  Curves are shown
for constant cloud mass $M(R_{\rm CO}) = 1\times 10^{6}$
$M_{\odot}$, $G_0'=10$,  and 
metallicities $Z' = 1.9$ ({\em dashed curve}), 
$Z' = 0.5$ ({\em dotted curve}).  Clouds of higher $\bar A_V$
have less surface dark gas relative to the CO interiors.
\label{fig:avmean}}

\end{document}